\documentclass[twocolumns]{aa}  

\usepackage{graphicx}
\usepackage{txfonts}
\begin{document}

   \title{Impacts of pure shocks in the BHR71 bipolar outflow}

%   \subtitle{I. Overviewing the $\kappa$-mechanism}

   \author{A. Gusdorf
          \inst{1}
          \and
          D. Riquelme\inst{2}
          \and
          S. Anderl\inst{3}
          \and 
          J. Eisl\"offel\inst{4}          
          \and
          C. Codella\inst{5}
          \and
          A. I. G\'omez-Ruiz\inst{2}
          \and
          U. U. Graf\inst{6}
          \and
          L. E. Kristensen\inst{7}
          \and
          \mbox{S. Leurini\inst{2}}
          \and
          B. Parise\inst{8}
          \and
          M. A. Requena-Torres\inst{2}
          \and
          O. Ricken\inst{2}
	 \and
          R. G\"usten\inst{2}}

 	  \institute{LERMA, Observatoire de Paris, \'Ecole Normale Sup\'erieure, PSL Research University, CNRS, UMR 8112, F-75014, Paris, France; Sorbonne Universit\'es, UPMC Univ. Paris 6, UMR 8112, LERMA, F-75005, Paris, France; \email{antoine.gusdorf@lra.ens.fr}
         \and
         Max Planck Institut f\"ur Radioastronomie, Auf dem H\"ugel 69, 53121 Bonn, Germany
         \and
         Univ. Grenoble Alpes, IPAG, F-38000 Grenoble, France \\
         CNRS, IPAG, F-38000 Grenoble, France
         \and
         Th\"uringer Landessternwarte, Sternwarte 5, 07778, Tautenburg, Germany
         \and
         INAF, Osservatorio Astrofisico di Arcetri, Largo E. Fermi 5, 50125, Firenze, Italy
         \and
         KOSMA, I. Physikalisches Institut, Universit\"at zu K\"oln, Z\"ulpicher Str. 77, 50937 K\"oln, Germany
         \and
         Harvard-Smithsonian Center for Astrophysics, 60 Garden Street, Cambridge, MA 02138, USA
         \and
         School of Physics \& Astronomy, Cardiff University, Queens Buildings, The parade, Cardiff CF24 3AA, UK
         }

   \date{Received xxx; accepted xxx}

% \abstract{}{}{}{}{} 
% 5 {} token are mandatory

%%%%%%%%%%%%%%%%%%%%%%%%%%%%%%%%%%%%%%%%%%%%%%%%%%%%%%%%%%%%%%%%%%%%%% 
  \abstract
  {During the formation of a star, material is ejected along powerful jets that impact the ambient material. This outflow regulates star formation by e.g. inducing turbulence and heating the surrounding gas. Understanding the associated shocks is therefore essential to the study of star formation.}
  {We present comparisons of shock models with CO, H$_2$, and SiO observations in a \lq pure' shock position in the BHR71 bipolar outflow. These comparisons provide an insight into the shock and pre-shock characteristics, and allow us to understand the energetic and chemical feedback of star formation on Galactic scales.}
   {New CO ($J_{\rm up}$ = 16, 11, 7, 6, 4, 3) observations from the shocked regions with the SOFIA and APEX telescopes are presented and combined with earlier H$_2$ and SiO data (from the \textit{Spitzer} and APEX telescopes). The integrated intensities are compared to a grid of models that were obtained from a magneto-hydrodynamical shock code which calculates the dynamical and chemical structure of these regions combined with a radiative transfer module based on the \lq large velocity gradient' approximation.}
   {The CO emission leads us to update the conclusions of our previous shock analysis: pre-shock densities of 10$^4$~cm$^{-3}$ and shock velocities around 20--25~km~s$^{-1}$ are still constrained, but older ages are inferred ($\sim$4000~years).}
%  % conclusions heading (optional), leave it empty if necessary 
   {We evaluate the contribution of shocks to the excitation of CO around forming stars. The SiO observations are compatible with a scenario where less than 4\% of the pre-shock SiO belongs to the grain mantles. We infer outflow parameters: a mass of $1.8\times10^{-2}~M_\odot$ was measured in our beam, in which a momentum of $0.4~M_\odot$~km~s$^{-1}$ is dissipated, corresponding to an energy of $4.2 \times 10^{43}$~erg. We analyse the energetics of the outflow species by species. Comparing our results with previous studies highlights their dependence on the method: H$_2$ observations only are not sufficient to evaluate the mass of outflows.}
%%%%%%%%%%%%%%%%%%%%%%%%%%%%%%%%%%%%%%%%%%%%%%%%%%%%%%%%%%%%%%%%%%%%%% 

   \keywords{Stars: formation --
                ISM: jets and outflows --
                ISM: individual objects: BHR71 --
                Submillimeter: ISM --
                Infrared: ISM
               }

   \maketitle
%
%________________________________________________________________

\section{Introduction}
\label{sec:intro}

Molecular shocks are ubiquitous in the interstellar medium (ISM) of our Galaxy. They can be associated with the formation of \lq ridges' at the convergence region of molecular clouds (e.g. W43, \citealt{Nguyenluong13}), to the jets and outflow systems related to the formation of low- to high-mass stars (from low, e.g. \citealt{Gomezruiz13}, to high, \citealt{Leurini13}, through intermediate, \citealt{Gomezruiz12} for examples; also see \citealt{Arce06}, \citealt{Frank14}, or \citealt{Tan14} for reviews), or to supernova remnants (SNRs) interacting with molecular clouds (e.g. W44, \citealt{Anderl14}). However, an analysis of these shocks is challenging because several physical processes are contributing to the observed emission. The W43 ridges are illuminated by an energetic UV radiation field coming from neighbouring \ion{H}{II} regions of star clusters (\citealt{Bally10}). Studying young stellar objects (YSO) by pointing on the central protostar leads in practice to a study of the combination of ejection shocks (that can be multiple, e.g. \citealt{Kristensen13}), infall processes, and UV illumination (the latter being all the more important when the mass of the forming object is large, e.g. \citealt{Visser12,Sanjosegarcia13}). Outflows from massive star-forming regions are also illuminated by strong radiation in the X-ray regime (e.g. W28~A2, \citealt{Rowell10}). Similarly, old SNRs are also often subject to X-rays, and also to $\gamma$-ray emission which are the signature of the acceleration of particles that locally took place shortly after the supernova explosion (e.g. \citealt{Hewitt09}). 

\lq Pure' molecular shock regions can be defined as regions where the physics and chemistry are dominantly driven by shocks. Studying these regions, therefore, is of crucial importance to investigating the feedback they exert on their environment, whether this feedback is energetic or chemical. From the energetic point of view, these studies allow us to assess the contribution of shocks to the excitation of e.g. CO on galactic scales, as observed by \textit{Herschel} (in NGC1068:~\citealt{Haileydunsheath12}, M82:~\citealt{Kamenetzky12}, NGC6240 and Mrk231:~\citealt{Meijerink13}, or NGC253:~\citealt{Rosenberg141} and Arp299:~\citealt{Rosenberg142}) from the inside of the Galaxy. From the chemical point of view, these studies allow us to investigate the formation paths of water (e.g. \citealt{Leurini141}) or more complex molecules (e.g. complex organic ones, \citealt{Belloche13}), and to understand their presence in planetary systems, for instance. These \lq pure' shock regions are not numerous. The most remarkable and well-studied pure shock region is the B1 knot of the L1157 bipolar outflow. Sufficiently distant from the central protostar, this region remains uncontaminated by infall or irradiation processes, and has been the subject of a number of studies dedicated to studying its energetics or chemical composition (\citealt{Gusdorf082, Codella10, Flower12, Benedettini13, Busquet14, Podio14}). In particular, L1157 was mapped by the GREAT (german receiver for astronomy at terahertz frequencies) receiver onboard SOFIA (Stratospheric Observatory for Infrared Astronomy) in CO (11--10), but the low signal-to-noise ratio prevented \citet{Eisloeffel12} from a thorough study of its energetics or chemistry. This article focus on the analysis of a similar shock position in the BHR71 outflow, the \lq southern twin' of L1157. Because of its southern location, BHR71 is an ideal target to be observed with ALMA (Atacama Large Millimeter/submillimeter Array), which will never be done for L1157 because of its high northern location. This article is organised as follows:~section~\ref{sec:tbhr71bo} presents our observations. Section~\ref{sec:restcod} presents the new CO data we made use of in our analysis, while the existing H$_2$ and SiO data is described in section~\ref{sec:rhed}. The results of shock modelling and their further use is exposed in section~\ref{sec:discuss}, and section~\ref{sec:conc} summarises our findings.  

\section{The BHR71 bipolar outflow}
\label{sec:tbhr71bo}

\subsection{Previous work}
\label{sub:smcoo}

The double bipolar outflow BHR71 (\citealt{Bourke97, Myers98, Parise06}) lies close to the plane of the sky. It is powered by two sources IRS1 and IRS2, separated by $\sim$3400 AU (\citealt{Bourke01}), of luminosities 13.5 and 0.5~$L_\odot$ (\citealt{Chen08}), relatively nearby ($\sim$200~pc, \citealt{Bourke95}). The present work builds on previous studies of the BHR71 bipolar outflow in H$_2$ (\citealt{Neufeld09} and \citealt{Giannini11}, hereafter N09 and Gia11), and H$_2$ plus SiO (\citealt{Gusdorf11}, hereafter G11). In their work, N09 mostly described the \textit{Spitzer} observations of the outflow. The InfraRed Spectrograph (IRS) was used to map the inner part of the outflow in the pure rotational transitions of H$_2$ as well as in \ion{Fe}{II} and \ion{S}{I} transitions. A region corresponding to approximately half the length of the outflow was covered by these observations around the driving protostars. The results were compared to previous observations of the entire region by the InfraRed Array Camera (IRAC) onboard the same telescope, showing that 30\% and 100\% of the luminosity of bands 3 and 4 could be accounted for by H$_2$ lines emission, similar to L1157. The pure rotational H$_2$ luminosity of the flow was estimated to be 4.4$\times10^{-2}$~$L_\odot$, less than 1/3 of that of L1157, but measured only from the fraction of the BHR71 outflow that was mapped. The H$_2$ mass above 100~K was constrained to be around 2.5$\times10^{-3}$~$M_\odot$, 20 times less than in L1157. However the H$_2$ densities for both outflows were constrained to $\sim$10$^{3.8}$~cm$^{-3}$. In Gia11, these IRS observations of H$_2$ were detailed line by line. An average column density of H$_2$ of around 10$^{20}$~cm$^{-2}$ was extracted from the map. Two temperature components were apparent, at $\sim$300 and $\sim$1500~K. A non-local thermodynamical equilibrium analysis was performed and yielded a high H$_2$ average density of a few 10$^6$~cm$^{-3}$. The total H$_2$ luminosity of the flow was estimated to be twice as high as the pure rotational lines, also based on previous observations of the rovibrational lines emission presented in \citet{Giannini04}. In a few positions where this was possible, the observations of pure rotational H$_2$ lines were combined with those of rovibrational lines to generate a more complete excitation diagram that was also compared to shock models, confirming the high value of H$_2$ density. 

In G11, the authors presented a shock-model analysis of various positions in the northern lobe of the BHR71 outflow. They used \textit{Spitzer} observations of H$_2$, and APEX observations of SiO to constrain shock models. Their analysis was hampered by the rather high beam-size of their SiO observations. However, they identified two positions where a shock analysis was favourable: the \lq SiO peak' and \lq SiO knot'. These positions lie at the apex of the inner bipolar structure of the outflow, relatively un-contaminated by envelope infall. They are far away from the protostars, and separated from them by the inner outflow cavity, hence as shielded as possible from their potential UV radiation. For all these reasons they are reminiscent of the L1157-B1 position, and are considered in the present study as \lq pure shock' positions.

\begin{figure*}[t]
   \centering
   \includegraphics[width=\textwidth]{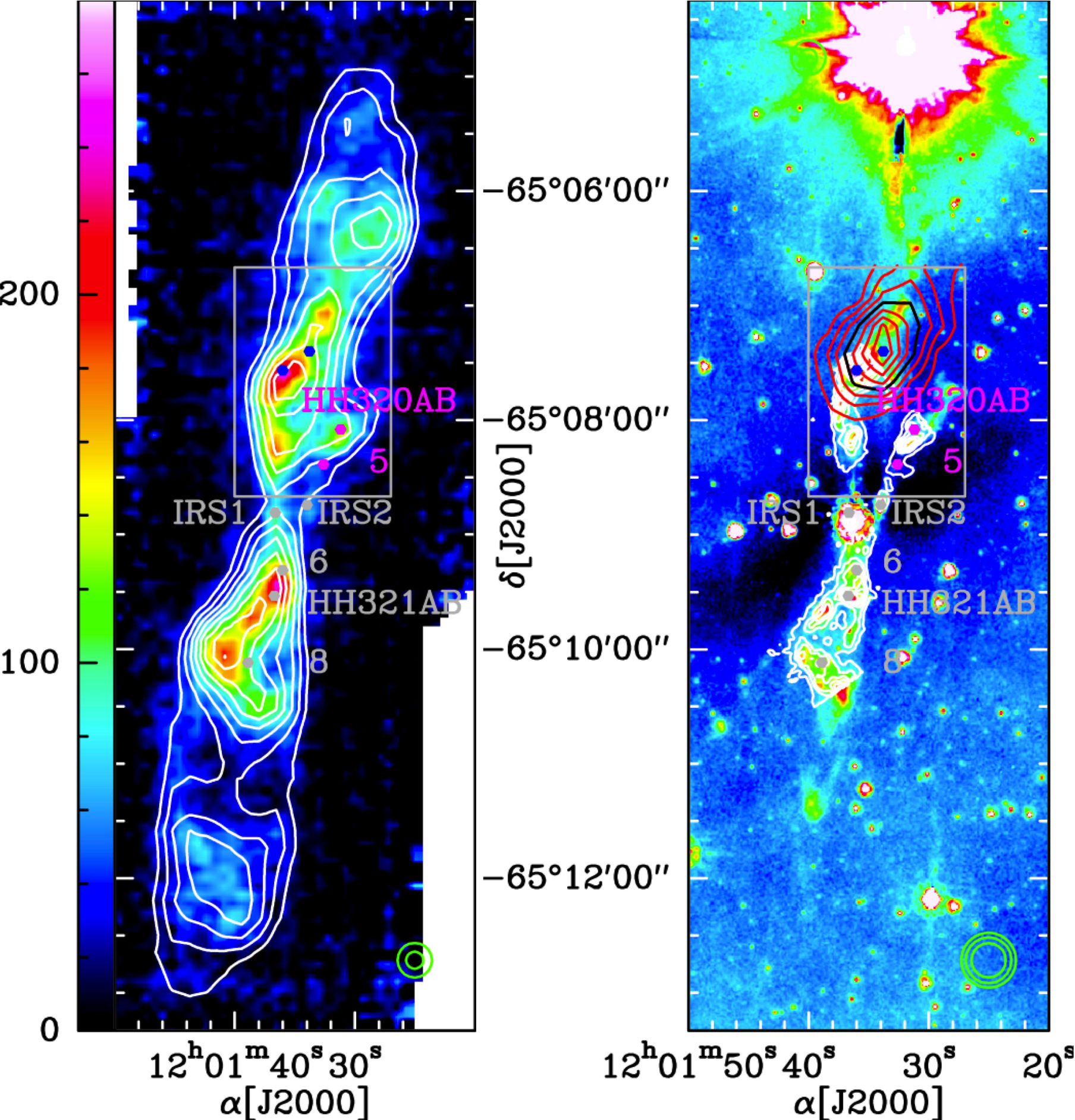}
      \caption{The BHR71 bipolar outflow in its entirety, in \textit{Left:} CO (6--5) (colours, associated with the wedge, in main beam temperature units, K km~s$^{-1}$) and (3--2) (white contours, from 30\% to 100\% of the signal in steps of 10\%), both observed by the APEX telescope and integrated between -50 and +50~km~s$^{-1}$, with the resolutions indicated by green circles in the lower right corner; \textit{Right:} the 8~$\mu$m emission detected by the \textit{Spitzer}/IRAC receiver (colours, N09), with the H$_2$ 0--0 S(5) emission as observed by the \textit{Spitzer}/IRS receiver in the inner parts of the outflow (white contours, from G11), and the SiO (5--4) emission (red and black contours, G11) in the upper lobe (the green circles in the lower right corner show the respective resolutions of CO (3--2), (11--10), and SiO (5--4)). On both maps, the grey inset is the field shown in figure~\ref{figure2}, the SiO peak and knot positions are indicated in blue, the knot 5 and HH320 region are in pink, and the IRS 1\&2, knots 6, 8, and HH321 region are indicated by grey dots.
              }
         \label{figure1}
   \end{figure*}

Figure~\ref{figure1} shows the entire BHR71 outflow as seen through these various datasets and highlights the particular positions as first introduced by \citet{Giannini04}: the so-called knots 5, 6, and 8 are thus indicated, as well as the two Herbig-Haro objects HH320A/B and HH321A/B. The driving protostars IRS 1 and IRS 2 are also marked. The two extra positions used in G11, \lq SiO peak' (of emission) and neighbouring \lq SiO knot' are also shown at the apex of the upper inner lobe of the outflow. In this figure, the previous and most recent observations can be compared (APEX, see section~\ref{sub:smcoo}):~intensity in the CO~(6--5) (colours) and (3--2) (white contours) transitions integrated between -50 and 50 km~s$^{-1}$ in the left panel, and \textit{Spitzer} IRAC (8~$\mu$m, colours) and IRS (H$_2$ 0--0 S(5), white contours) maps overlaid with SiO (5--4) map (red contours in the upper lobe) in the right panel. The white contours then show how much of the outflow was observed by the IRS onboard \textit{Spitzer}.

\subsection{APEX observations of CO}
\label{sub:smcoo}

APEX\footnote{This publication is partly based on data acquired with the Atacama Pathfinder EXperiment (APEX). APEX is a collaboration between the Max-Planck-Institut f\"ur Radioastronomie (MPIfR), the European Southern Observatory, and the Onsala Space Observatory.} observations of the entire BHR71 outflow were conducted in 2013 and 2014. The analysis of the whole maps is out of the scope of the present work, and will be the subject of a forthcoming publication. In the present study, we used information inferred from the full maps in the $^{13}$CO (3--2), $^{12}$CO (3--2), (4--3), (6--5), and (7--6) transitions, which we briefly describe. The APEX observations towards BHR71 were conducted on several days:~June 3 and 4, 2013, and June 28, 2014. We used the heterodyne receivers FLASH345, FLASH460 (First Light APEX Submillimeter Heterodyne receiver, \citealt{Heyminck06}), and CHAMP$^+$ (Carbon Heterodyne Array of the MPIfR, \citealt{Kasemann06,Guesten08}), in combination with the MPIfR fast Fourier transform spectrometer backend (XFFTS, \citealt{Klein12}). The central position of all observations was $\alpha_{[\rm{J}2000]}$=$12^h01^m36\fs3$, $\delta_{[\rm{J}2000]}$=$-65^\circ08'53$\farcs$0$. We checked the focus at the beginning of each observing session, after sunrise and sunset on Saturn. We checked line and continuum pointing locally on IRC+10216, 07454-7112, or {\bf $\eta$ Car}. The pointing accuracy was better than $\sim$5$''$ rms, regardless of which receiver we used. Table~\ref{table1} contains the main characteristics of the observed lines and corresponding observing set-ups. The observations were performed in position-switching/on-the-fly mode using the APECS software \citep{Muders06}. We reduced the data with the CLASS software (see http://www.iram.fr/IRAMFR/GILDAS). This reduction included baseline subtraction, spatial, and spectral regridding. For all observations, the maximum number of channels available in the backend was used (8192). We obtained maps for all considered transitions, covering the field of the whole outflow shown in figure~\ref{figure1}. 

\begin{figure}
   \centering
   \includegraphics[width=9cm]{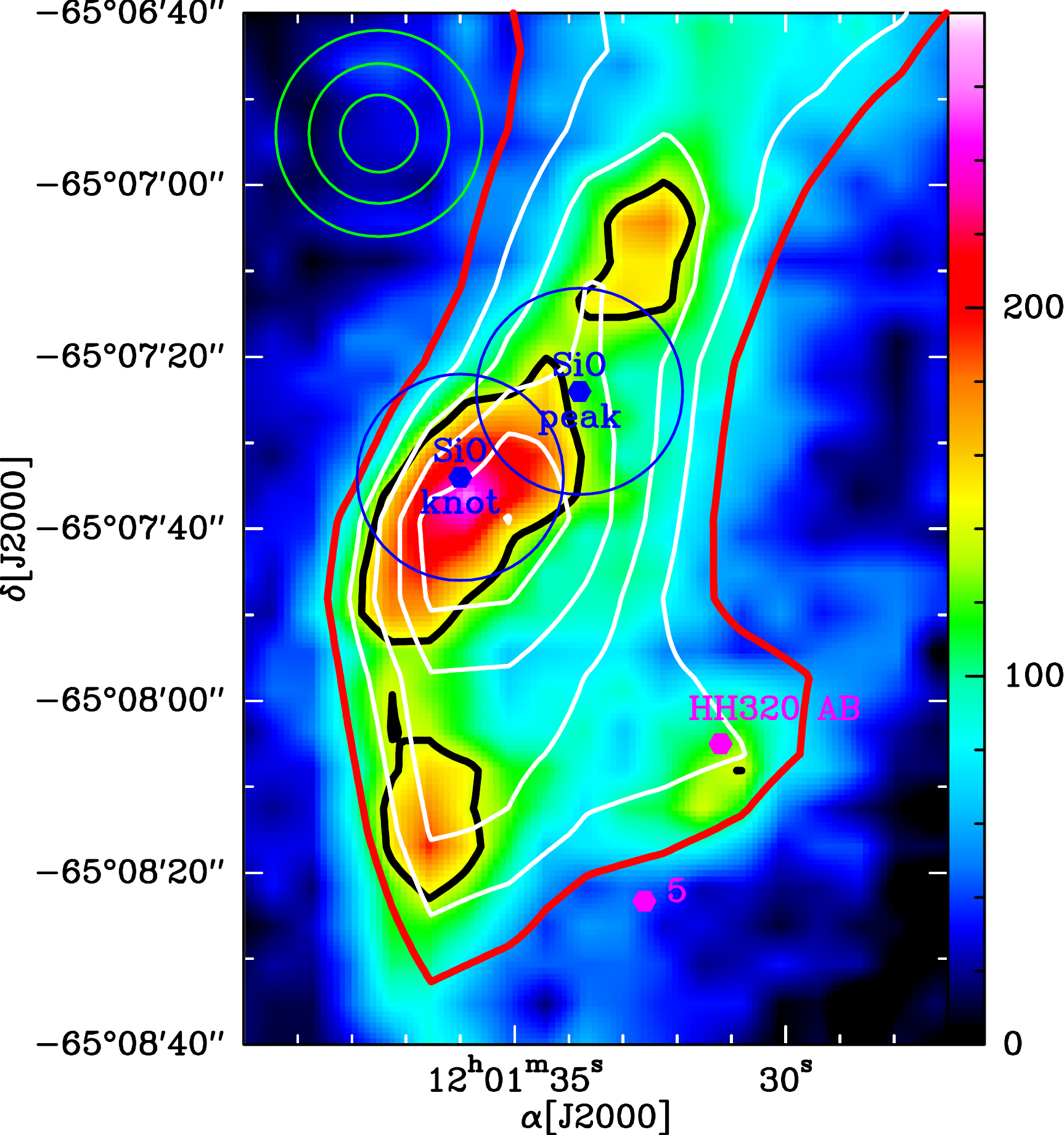}
      \caption{Overlay of the velocity-integrated CO (6--5) (colour background) with the CO (3--2) (white contours) emission observed with the APEX telescope in the inner upper lobe of the BHR71 outflow. For both lines, the intensity was integrated between -50 and 50 km s$^{-1}$. The wedge unit is K km s$^{-1}$ in main beam temperature. The CO (3--2) contours are from 50 to 100\% of the maximum, in steps of 10\%. The half-maximum contours of the CO (3--2) and (6--5) maps are indicated in red and black, respectively. The dark blue circles indicate the positions and beam size of the SOFIA/GREAT observations for the CO (11--10) transition. The APEX and SOFIA beam sizes of our CO (6--5), (16--15) and (11--10) observations are also provided (upper left corner light green circles, see also table~\ref{table1}). The pink hexagons mark the position of the so-called knot 5 and HH320AB object.}
         \label{figure2}
   \end{figure}

Figure~\ref{figure2} shows the velocity-integrated CO (6--5) broad-line emission (colours, resolution 9$''$) in the upper inner part of BHR71, overlaid with the CO (3--2) (white contours, resolution 18$''$). The higher angular resolution of the CO (6--5) line emission shows that it traces the walls of the cavity of the outflow associated with IRS1 (see Figure~\ref{figure1}), with a local maximum of emission also corresponding to the outflow driven by IRS2 (the HH320 AB position in the map). The outflow is seen close to the plane of the sky. The two positions of interest identified by G11 are marked by a blue hexagon and circle corresponding to the beam of our CO (11--10) observations with GREAT (which will be the focus of our analysis, see sections~\ref{sub:fircos} and \ref{sec:restcod}). The position of the SiO peak, detected through 28$''$-resolution observations of the SiO (5--4) is localised between two emission peaks in CO (6--5). The most prominent of these CO peaks is the southern one, which corresponds to the so-called SiO knot;~it coincides with an H$_2$ emission peak (also see figure~\ref{figure6}, which overlays the same CO (6--5) field with the H$_2$ 0--0 S(5) data of N09 and Gia11).  The half-maximum emission contours are also given in this map in thick black and red contours for the (6--5) and (3--2) lines, respectively, at their nominal resolutions.
     
 \begin{table*}
\caption{Observed lines and corresponding telescope parameters for the APEX and SOFIA observations of BHR71.}             
\label{table1}      
\centering                          
\begin{tabular}{l  c  c  c  c c c c }        
\hline           
species & $^{12}$CO & $^{12}$CO & $^{12}$CO & $^{12}$CO & $^{12}$CO & $^{12}$CO & \multicolumn{1}{c}{$^{13}$CO} \\
\hline
line & (3--2) & (4--3) & (6--5) & (7--6) & (11--10) & (16--15) & (3--2)  \\
telescope & APEX & APEX & APEX & APEX & SOFIA & SOFIA & APEX \\
\hline
\hline
\footnotesize{$\nu$ (GHz)} & \footnotesize{345.796} & \footnotesize{461.041} & \footnotesize{691.473} & \footnotesize{806.652} & \footnotesize{1267.014} & \footnotesize{1841.346} & \footnotesize{330.588} \\
\footnotesize{FWHM ($''$)} & \footnotesize{18.1} & \footnotesize{13.5} & \footnotesize{9.0} & \footnotesize{7.7} & \footnotesize{24} & \footnotesize{16.3} & \footnotesize{18.9}  \\
\footnotesize{sampling ($''$)} & \footnotesize{6} & \footnotesize{6} & \footnotesize{4} & \footnotesize{4} & \footnotesize{pointed} & \footnotesize{pointed} & \footnotesize{6} \\
\hline
\footnotesize{receiver} & \footnotesize{FLASH345} & \footnotesize{FLASH460} & \footnotesize{CHAMP$^+$} & \footnotesize{CHAMP$^+$} & \footnotesize{GREAT} & \footnotesize{GREAT} & \footnotesize{FLASH345}  \\
\footnotesize{observing days} & \footnotesize{2013-06-03} & \footnotesize{2013-06-03} & \footnotesize{2013-06-04} & \footnotesize{2013-06-04} & \footnotesize{2013-07-23} & \footnotesize{2013-07-23} & \footnotesize{2013-06-03} \\
 & \footnotesize{2013-06-04} & \footnotesize{2013-06-04} & \footnotesize{2014-06-28} & \footnotesize{2014-06-28} & \footnotesize{2013-07-23} & \footnotesize{2013-07-23} & \footnotesize{2013-06-04} \\
\hline
\footnotesize{$F_{ \rm eff}$} & 0.95 &0.95 & 0.95 & 0.95 & \footnotesize{0.97} & \footnotesize{0.97} & 0.95 \\
\footnotesize{$B_{ \rm eff}$} & 0.69 &0.60 & 0.42 & 0.35 & \footnotesize{0.67} & \footnotesize{0.67} & 0.69 \\
\hline 
\footnotesize{$T_{\rm sys}$ (K)} & 169-193 & 408-610 & 3650-10000 & 12000-34000 & \footnotesize{3274--3307} & \footnotesize{3160--3207} & 182-224  \\
\footnotesize{$\Delta \varv$ (km s$^{-1}$)} & \footnotesize{0.331} & \footnotesize{0.496} & \footnotesize{0.635} & \footnotesize{0.544} & \footnotesize{0.018} & \footnotesize{0.012} & \footnotesize{0.346}  \\    
\hline                                  
\footnotesize{reference offset ($''$)} & \footnotesize{(-120,260)} & \footnotesize{(-120,260)} & \footnotesize{(-120,260)} & \footnotesize{(-120,260)} & \footnotesize{beamswitch} & \footnotesize{beamswitch} & \footnotesize{(-120,260)} \\
\hline
\end{tabular}
\end{table*}

\subsection{SOFIA-GREAT observations of CO}   
\label{sub:fircos}

The observations towards BHR71 were conducted with the GREAT\footnote{GREAT is a development by the MPI f\"ur Radioastronomie and the KOSMA$/$Universit\"at zu K\"oln (K\"olner Observatorium f\"ur SubMillimeter Astronomie), in cooperation with the Max-Planck-Institut f\"ur Sonnensystemforschung and the Deutsches Zentrum f\"ur Luft- und Raumfahrt Institut f\"ur Planetenforschung.} spectrometer (\citealt{Heyminck12}) during SOFIA's Cycle 1 \lq southern deployment' flight on July 23rd, 2013. We observed two positions, towards the northern apex of the inner outflow structure: the SiO knot and peak, as defined in G11 (figures~\ref{figure1} and \ref{figure2}). We tuned the receiver to the CO (11--10) and (16--15) lines frequency 1267.014 GHz LSB and 1841.346 GHz USB. We connected the receiver to a digital FFT spectrometer (\citealt{Klein12}) providing a bandwidth of 1.5 GHz with respective spectral resolutions of 0.018 and 0.012~km~s$^{-1}$. The observations were performed in double beam-switching mode, with an amplitude of 80$''$ (or a throw of 160$''$) at the position angle of 135$^\circ$ (NE--SW) and a phase time of 0.5 sec. The nominal focus position was updated regularly against temperature drifts of the telescope structure. The pointing was established with the optical guide cameras to an accuracy of $\sim$5$''$. The line and observation parameters are listed in table~\ref{table1}. The integration time was 13 min ON source for each line, for respective final r.m.s of $\sim$0.70 and 0.75~K. The data were calibrated with the KOSMA/GREAT calibrator (\citealt{Guan12}), removing residual telluric lines, and subsequently processed with the CLASS software\footnote{http://www.iram.fr/IRAMFR/GILDAS}.
      
\section{Results: the CO data}   
\label{sec:restcod}   

\subsection{Spectra}   
\label{sub:spe}   

   \begin{figure*}
   \centering
   \includegraphics[width=\textwidth]{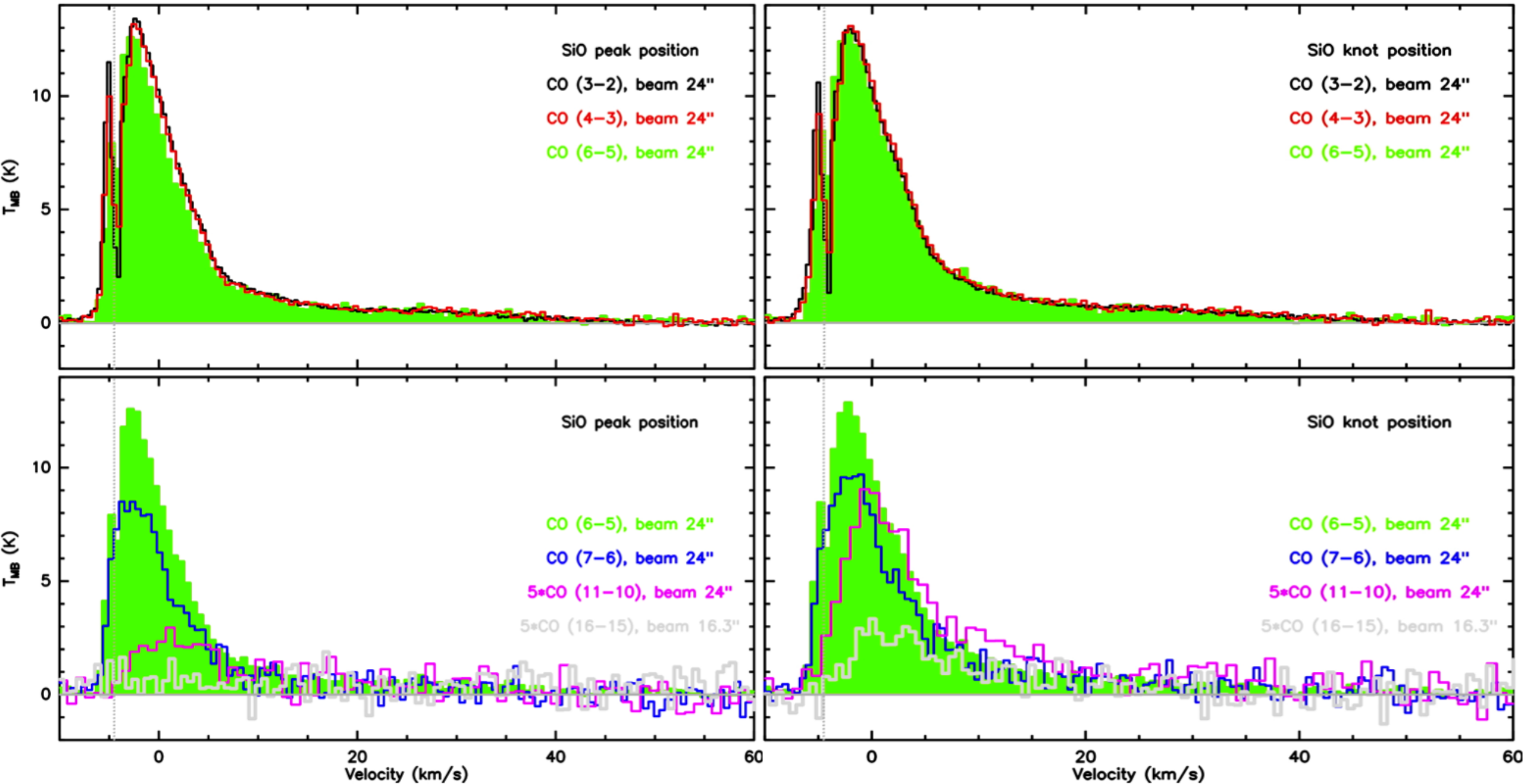} 
      \caption{CO transitions in the SiO peak (left panels) and knot (right panels) positions indicated in figure~\ref{figure2}: APEX (3--2), black line~; (4--3), red line~; (6--5), green line and histograms (upper panels); (7--6), dark blue lines~; SOFIA (11--10), pink lines and (16--15), grey lines (lower panels, overlaid on the green histograms of the 6--5 transition). The last two were multiplied by five for comparison purposes. Respective spectral resolutions are 0.33, 0.50, 0.64, 0.54, 0.90 and 0.62~km~s$^{-1}$. The vertical dotted line marks the cloud velocity, -4.5~km~s$^{-1}$ \citep{Bourke952}.    
              }
         \label{figure3}
   \end{figure*}   
   
Figure~\ref{figure3} presents all $^{12}$CO spectra obtained in the two targeted positions in the course of our APEX ($J_{\rm up}$ = 3,4,6,7) and SOFIA ($J_{\rm up}$ = 11, 16) observations, after convolution to the same angular resolution, that of the CO (11--10) transition (see table~\ref{table1} for observing parameters related to each of these lines). The only exception to this convolution is the CO (16--15) line, for which only a single-point observations is available. As the line was observed at an angular resolution of roughly 16\farcs3, we could in principle use the integrated intensity extracted from this line as an upper limit to that convolved to the 24$''$ resolution. In fact, we corrected this value for the beam dilution effect in our analysis, see discussion in section~\ref{sub:csm} on how we made use of this line. On the figure, we split the spectra in two groups for visibility purposes ($J_{\rm up}$= 3,4,6 in the upper panels, and $J_{\rm up}$= 6, 7, 11, 16 in the lower panels).

In both positions, all lines exhibit a profile typical of the presence of a pure shock, with wings extending towards red-shifted velocities, up to 30--40~km~s$^{-1}$; up to $J_{\rm up} = 6$, self-absorption is also detected at the velocity of the cloud (around \mbox{$-$4.5 km s$^{-1}$}). The CO~(3--2) and (4--3) profiles coincide very well, suggesting that these lines are optically thick. This is confirmed for the CO (3--2) line, for which we observed the $^{13}$CO isotopologue on both positions (see section~\ref{sub:co32opa}, table~\ref{table1} and figure~\ref{figure5} for the spectra obtained on both positions). In both positions, the CO~(6--5) profiles start to differentiate from the lower-lying profiles. This departure is confirmed in the higher-lying transitions, and reveals that the lines from $J_{\rm up} = 6$ are probably excited in different layers of the shocked region than their lower-lying counterparts. Globally, it can be seen that the excitation conditions vary with the position:~the $J_{\rm up}$ = 11 and 16 profiles differ, and the \mbox{CO~(16--15)} is not even detected in the SiO peak position. For both positions, the excitation conditions also vary with the velocity: close to the systemic velocity, the lower-$J$ line emission greatly dominates, an effect previously reported in L1157~B1 by~\citet{Lefloch12}. Because of the associated weak or absent detection in the SOFIA lines, and because the two selected positions are not independent, we decided to exclude the SiO peak from the shock analysis presented in section~\ref{sec:discuss} (also see an additional argument in section~\ref{sub:h2o}).

\subsection{CO (3--2) line opacities}
\label{sub:co32opa}

\begin{figure}
   \centering
   \includegraphics[width=9cm]{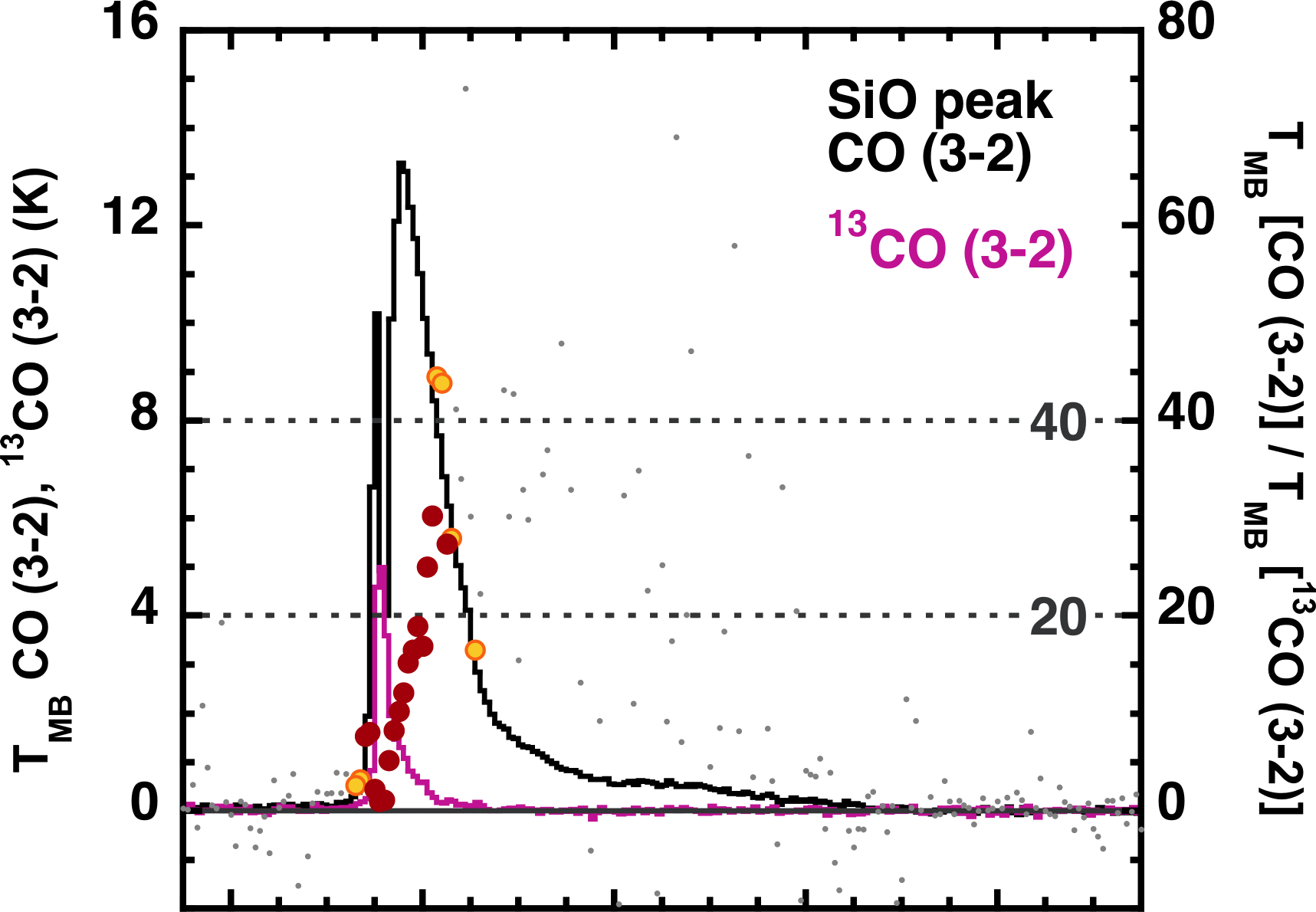}
   \includegraphics[width=9cm]{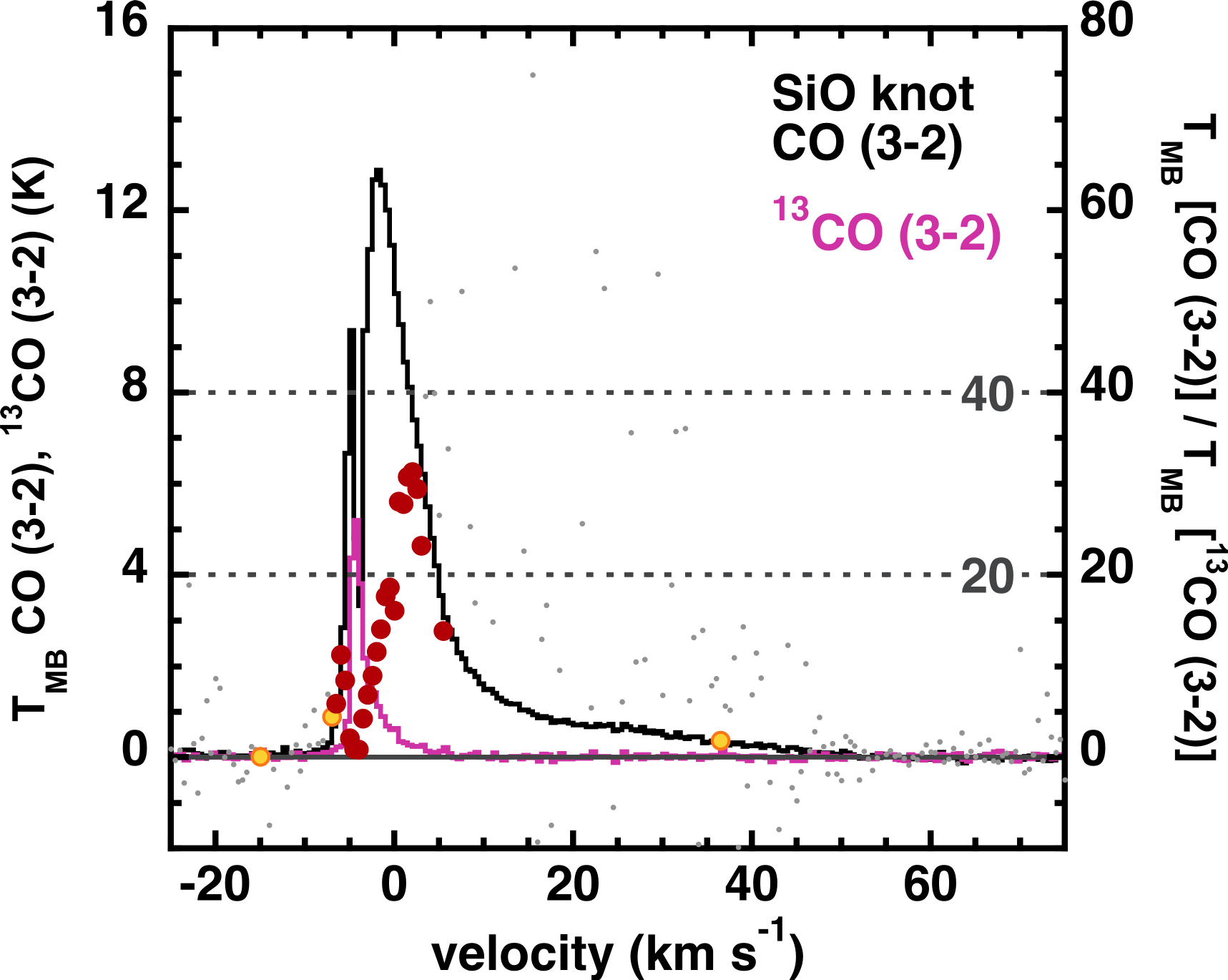}
      \caption{Comparison of the $^{12}$CO (black line) and $^{13}$CO (pink line) (3--2) emission profiles obtained in the SiO peak (upper panel) and knot (lower panel) positions. The ratio of main-beam temperatures is also shown in the form of red dots (for the channels where the $^{13}$CO detection is over 3$\sigma$), orange dots (for the channels where the $^{13}$CO detection is between 2 and 3$\sigma$), and grey dots (for the remaining channels). Associated with this distribution are the grey horizontal dotted lines, which show values of  20 and 40 for this ratio.}
         \label{figure5}
   \end{figure}

Figure~\ref{figure5} shows our spectra of the $^{13}$CO and $^{12}$CO lines in the SiO peak and knot positions at the same spectral and spatial resolutions. These spectra allow us to plot the ratio of the line temperature. For each velocity channel, the line temperature ratio $^{12}$CO / $^{13}$CO is hence also shown (right ordinates) with the following colour-code:~red dots show the ratios for the velocity channels where the $^{13}$CO is detected at more than 3$\sigma$, and orange dots where the $^{13}$CO detection was obtained with a confidence level between 2 and 3$\sigma$. The grey dots are for all the other channels. Horizontal dashed lines show the reference values of 20 and 40 for these ratios. The orange and red dots all lie below 30. Assuming an identical excitation temperature for the $^{12}$CO and $^{13}$CO lines, and a typical interstellar abundance ratio of 50--60 (e.g. \citealt{Langer93}), these line ratios yield minimum opacity values of 3 for the $^{12}$CO lines. We therefore conclude that the emission is optically thick at least in the low-velocity regime of the spectral wings. The large optical thickness is consistent with the constancy of the line integrated intensities (in the non-absorbed components) from CO (3--2) and (4--3), when convolved to the same resolution(s). As our observations yield a minimum optical thickness of 3 up in the wings, we decided to exclude both the CO~(3--2) and (4--3) lines from our analysis.

\subsection{Dynamical age of the outflow}
\label{sub:daoto}   

An important parameter for the modelling of young outflows is their age. As we have mapped the entire outflow, we are able to give an upper limit of its age, simply obtained by dividing the distance between the furthest point from the driving protostar and the protostar by the associated linewidth. The positions we consider belong to the northern lobe of the outflow powered by the IRS 1 protostar. As the furthest point with significant CO emission belonging to this outflow, we adopted the furthest point on the 10\% CO (3--2) emission contour or the furthest point with 3$\sigma$ emission away from the driving protostar, and found that these two approaches were equivalent. Indeed, the offset from both points relative to the IRS1 position is about \mbox{($-$72$''$, 280$''$)}, corresponding to a distance of $\sim$289$''$. At a distance of 200~pc, this translates in $\sim$5.8$\times 10^{4}$~AU. In this position, the full-width at zero intensity of both the CO (3--2) and (6--5) (unshown) lines is about 15--20 km~s$^{-1}$, which converts to a dynamical age of (1.4--1.8)$\times 10^4$~years. Of course, this is an upper limit of the real age of the outflow, as it is likely that the apex of the outflow was associated with greater velocities in the past. A similar measurement for the SiO knot position (i.e. dividing its distance to the protostar by the full-width at zero intensity) yields a dynamical age of $\sim$1800 years.

%As the furthest point with significant CO emission belonging to this outflow, we adopted the furthest point on the furthest white contour in the left panel of figure~\ref{figure1}. Its offset relative to the IRS1 position is \mbox{($-$50$''$, 231$''$)}, corresponding to a distance of $\sim$236$''$. At a distance of 200~pc, this translates in $\sim$4.7$\times 10^{4}$~AU. In this position, the full width at zero intensity of both the CO (3--2) and (6--5) (unshown) lines is about 15 km~s$^{-1}$ from the systemic velocity, which converts to a dynamical age of 1.5$\times 10^4$~years. Of course, this is an upper limit of the real age of the outflow, as it is likely that the apex of the outflow was associated with greater velocities in the past. A similar measurement for the SiO knot position (i.e. dividing its distance to the proto-star by the full width at zero intensity) yields a dynamical age of $\sim$1800 years.

\section{Results: existing data}
\label{sec:rhed}

\subsection{H$_2$ observations}
\label{sub:h2o}

   \begin{figure}
   \centering
   \includegraphics[width=9cm]{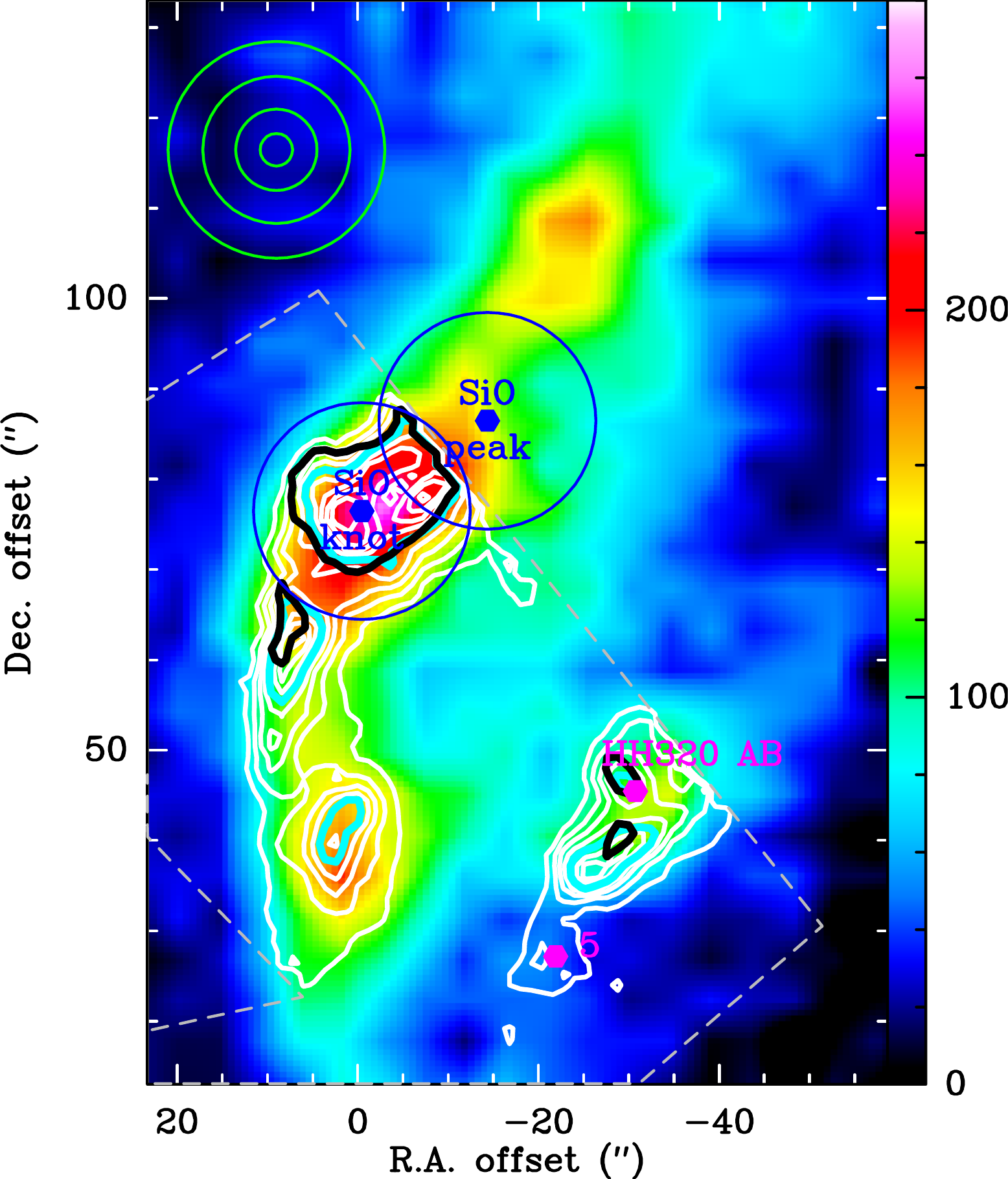}
      \caption{Overlay of the map of CO (6--5) emission observed by the APEX telescope (colour background) with the H$_2$ 0--0 S(5) emission (white contours), observed with the \textit{Spitzer} telescope. The wedge unit is K km s$^{-1}$ (main beam temperature) and refers to the CO observations. The H$_2$~0--0~S(5) contours are from 10\% to 100\%, in steps of 10\%. The light blue contour defines the half-maximum contour of this transition. Like in figure~\ref{figure2}, the blue circles and dots indicate the positions and beam sizes of the SOFIA/GREAT observations. The beam and pixel sizes of the CO (6--5) (11--10), (16--15), and H$_2$~0--0~S(5) observations are the green circles in the upper left corner. The black contour delineates the half-maximum contour of the H$_2$~0--0~S(2) transition. The field is the same as in figure~\ref{figure2}, and the knot 5 and HH320 region are in pink. The field covered by \textit{Spitzer}/IRS to observe the H$_2$ emission is indicated in dashed grey line. It excludes the SiO peak position.    
              }
         \label{figure6}
   \end{figure}
   
We used the \textit{Spitzer}/IRS observations of the H$_2$ pure rotational transitions, 0--0~S(0) up to S(7), reported and analysed in N09, Gia11, and G11. The reduced data kindly communicated to us by David Neufeld contain rotational transition maps, with 3\farcs6 angular resolution, centred on $\alpha_{[\rm{J}2000]}$=$12^h01^m36\fs31$, $\delta_{[\rm{J}2000]}$=$-65^\circ08'53$\farcs$02$. Figure~\ref{figure6} shows an overlay of our APEX CO~(6--5) map with the H$_2$~0--0~S(5) region observed by \textit{Spitzer}, both at their nominal resolution. The figure shows coinciding maxima between the two datasets in the selected position, and a slightly different emission distribution. This might be the effect of the better spatial resolution of the H$_2$ data, which reveal more peaks than in CO. This overlay also shows the similar morphology of the emission of the S(2) (half-maximum contour in black) and S(5) (half-maximum contour in light blue) transitions on the region we chose to analyse. Unfortunately, the figure also shows the coverage of the H$_2$ emission observations, and that the SiO peak position was not covered by the H$_2$ observations, which is another reason to exclude it from our shock analysis presented in section~\ref{sec:discuss}.

As part of our modelling (see Sect.~\ref{sec:discuss}), we used the excitation diagram derived for the selected emission region around the SiO knot position. The H$_2$ excitation diagram displays ln($N_{\varv j}/g_j$) as a function of $E_{\varv j}/k_{\rm B}$, where $N_{\varv j}$ (cm$^{-2}$) is the column density of the rovibrational level ($v, J$), $E_{\varv j}/k_{\rm B}$ is its excitation energy (in K), and $g_j = (2j+1)(2I+1)$ its statistical weight (with $I=1$ and $I=0$ in the respective cases of ortho- and para-H$_2$). If the gas is thermalised at a single temperature, all points in the diagram thus fall on a straight~line. 

The initial resolution of the maps is 3\farcs6, but we wanted to operate at the same resolution as for CO. Unfortunately, the SiO knot position lies at the edge of the field covered in the H$_2$ map, as can be seen in figure~\ref{figure6}. We consequently used three different methods for extracting the H$_2$ line fluxes. First, we extracted the flux from the initial map at its nominal 3\farcs6 resolution; second, we used the first method, but applied to a map that was convolved to the 24$''$ resolution; and third, we associated a filling factor of 0.2 to the fluxes obtained through the second method.

Because of the location of the SiO knot and the rather large beam of our analysis, the first two methods provide lower limits to the H$_2$ line fluxes. On the contrary, with method (3) we voluntarily extracted upper limits to these fluxes. We then used an average value between the values inferred from the three methods, and computed rather large errorbars based on the combination of all three methods. The values were corrected from interstellar extinction following the treatment already applied in G11. The result can be seen in the excitation diagram shown in the lower panel of figure~\ref{figure7}, where points corresponding to methods (1), (2), (3) are shown in black empty diamonds, upward and downward triangle, and the resulting average points are the black squares, respectively. The overall average H$_2$ values we used to build our figure are given in table~\ref{tablea1}.

\subsection{SiO}
\label{sub:sio}

As part of our study, we also decided to re-analyse the SiO observations already presented in G11. The emission from three lines was mapped in the northern lobe of the outflow:~SiO (5--4), (6--5), and (8--7), and the corresponding spectra were extracted in the SiO peak and knot positions. As maps were obtained, all spectra were convolved to the resolution of the SiO (5--4) line, about 28$''$. In the present study, we hence corrected the SiO dataset for the slight difference in resolution between this value and that of the CO (11--10) observations, 24$''$, by simply multiplying the G11 integrated intensities by (28/24)$^2$. Also, given the rather large beam size, we generated an integrated intensity diagram (displaying $\int T_{\rm MB} \rm{d}\varv$ against $J_{\rm up}$ for each transition) for three different filling factor values, 1, 0.5, and 0.2. The result can be seen in the form of  respective black, dark grey and light grey squares in figure~\ref{figure8}. The SiO values we used to build our figure are listed in table~\ref{tablea1}.

\section{Discussion}
\label{sec:discuss}

In the following we seek to constrain the physical conditions in the shocked regions. Given the large number of observational constraints, covering several species and several associated transitions, the method-of-choice for obtaining these physical conditions is a comparison to sophisticated shock models. The results of the shock modelling forms the foundation of the discussion: what characterises the chemistry of the outflow, in particular with respect to SiO, and what characterises the energetics. 

\subsection{Constraining shock models}
\label{sub:csm}

\begin{figure}
   \centering
   \includegraphics[width=9cm]{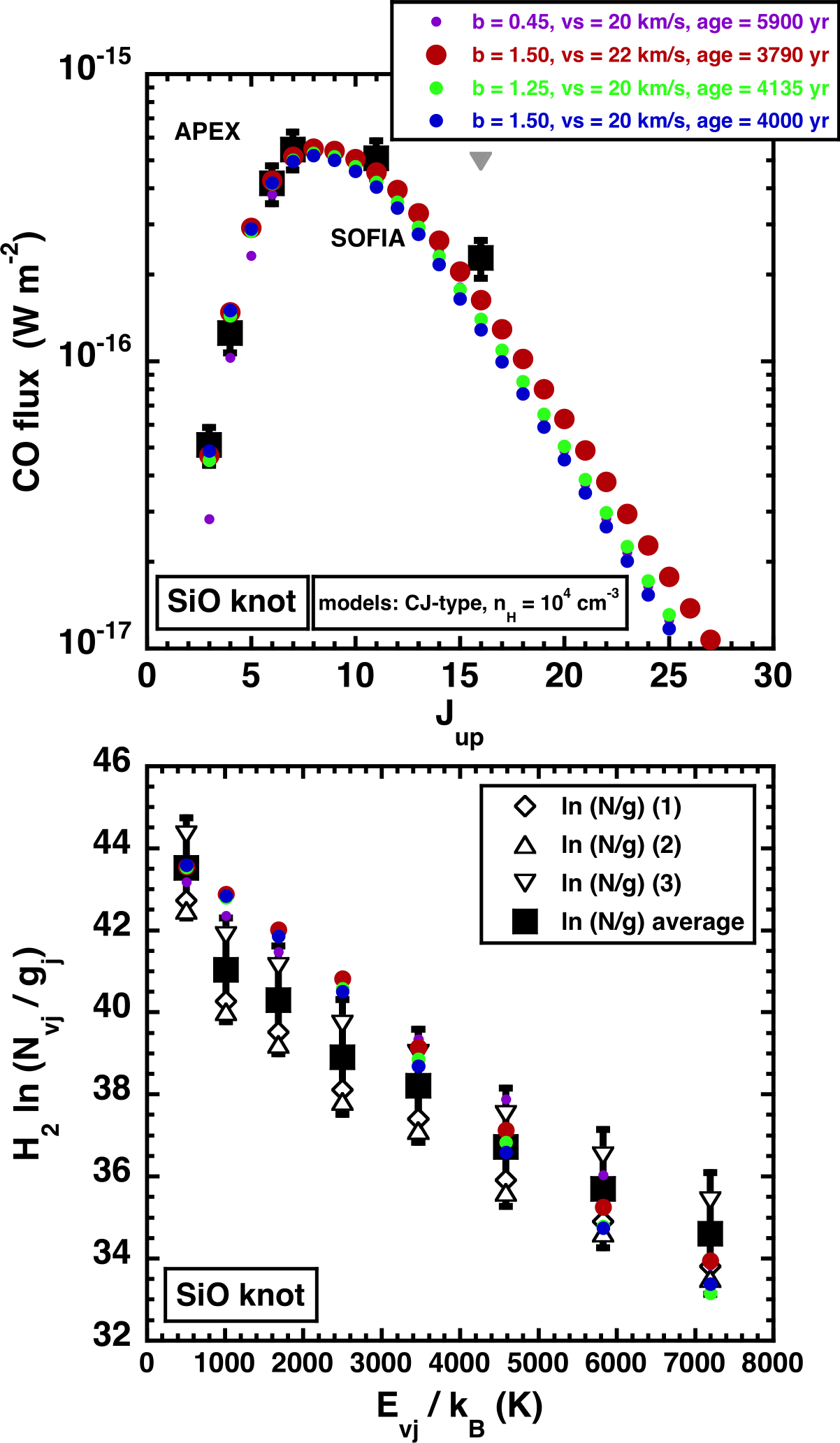}
      \caption{Our model-observations comparisons. \textit{Upper panel:}~CO flux diagram over a beam of 24$''$; the observations in the SiO knot position are the black squares, and the model results are the coloured circles (see text). The CO (16--15) observational point is corrected for beam-size effect. The uncorrected point is the upper limit (see text), indicated by the grey arrow in this panel. \textit{Lower panel:}~H$_2$ excitation diagram for the SiO knot position, extracted following the different procedures described in the text (empty symbols), global average in black squares, and model results in coloured circles (see text, with the same code as in the upper panel).}
         \label{figure7}
   \end{figure}

Shock models are used to constrain the physical conditions in the outflow shocks through comparison with H$_2$ and CO, following the methods already used in \citet{Gusdorf081, Gusdorf12} and \citet{Anderl14}. We generated CO flux diagrams and H$_2$ excitation diagrams from the observations of the SiO knot position, and compared them with results from a grid of models computed with the \lq Paris-Durham' (e.g. \citealt{Flower031}) 1-D shock model. The H$_2$ diagram has already been introduced in section~\ref{sub:h2o} and is shown in figure~\ref{figure7}. Concerning the observable associated with CO, we chose to display the data in the form of a flux diagram (line flux vs. $J_{\rm up}$), because of the increasing number of extragalactic studies that use this form. In the present case, we benefit from the fact that our observations were pointed towards a pure shock position. To generate a CO flux diagram, we consequently integrated the signal obtained over the whole line profile at the angular resolution of 24$''$, associated with a filling factor of 1 (see section~\ref{sec:ff} for an explanation on filling factors for various CO transitions). Because of their opacity, and their likeliness to be contaminated by ambient gas, we excluded the (3--2) lines from our analysis (see section~\ref{sub:co32opa}). Given the opacity values inferred from this line, we also excluded the (4--3) line from our fitting procedure. Overall, the two observational tools we used can be seen in figure~\ref{figure7}: a flux diagram for CO (upper panel) and an excitation diagram for H$_2$ (lower panel) on which the observational points for the SiO knot are always shown in black squares. The CO values we used to build our figures are listed in table~\ref{tablea1}. The observational flux diagram for the SiO peak position can be found in Section~\ref{sec:tcofditspp} (Figure~\ref{figurec1}).

%\begin{table}
%\caption{Input parameters of our grid of stationary C- and non-stationary CJ-type shock models.}             
%\label{table2}      
%\centering                          
%\begin{tabular}{c c c c}        
%\hline            
%type & $n_{\rm H}$ (cm$^{-3}$) & b & $\varv_{\rm s}$ (km s$^{-1}$)  \\    
%\hline \hline  
%C & 10$^3$ & 0.5,1,2 & \tiny{20,25,30,35} \\
%C & 10$^4$ & $\in$ [0.45 ; 2] & \tiny{10,15,20,22,25,28,30,32,35,55} \\
%C & 10$^5$ & $\in$ [0.30 ; 2] & \tiny{10,12,15,18,20,40} \\
%C & 10$^6$ & 1 & \tiny{10,12,15,18,20,22,25,27,30,32} \\
%\hline                %
%CJ & 10$^3$ & 0.5,1,2 & \tiny{20,25,30,35} \\
%CJ & 10$^4$ & $\in$ [0.45 ; 2] & \tiny{10,12,15,20,22, 25,28,30,32,35} \\
%CJ & 10$^5$ & $\in$ [0.30 ; 2] & \tiny{10,12,15,18,20} \\
%\hline
%\end{tabular}
%\end{table}

\begin{table*}
\caption{Shock model parameters. Grid intervals $\Delta x$ are listed as minimum and maximum increments found in the grid.}             
\label{table2}     
\centering                          
\begin{tabular}{l  c  c  c  c  c  c  c  c }        
\hline           
 Shock type & Number of models & $\varv$ [km s$^{-1}$] & $\Delta \varv$ [km s${}^{-1}$] & $b$ & $\Delta b$&  $n_{\rm H}$ [cm${}^{-3}$] & Age [yr] & $\Delta$Age  [yr]\\
\hline
\hline
            \noalign{\smallskip}
 C-type\tablefootmark{a} & 108 & 20 -- 55 &  2 -- 5&  0.45 -- 2 &0.15 -- 0.25 &10$^3$, 10$^4$, 10$^5$, 10$^6$ &   -- &  --\\
 C-type\tablefootmark{b}  &  9 &  20 -- 40 &  10 &  1 -- 3 &  0.5 &  10$^5$ & -- & --\\
 J-type &  21 & 10--50 & 2 --15 & 0.1 & -- & 10$^4$, 10$^5$ & -- & --\\
CJ-type &  1035 & 10 -- 50 & 2 --5 & 0.3 --2 &  0.15 -- 0.25 & 10$^3$, 10$^4$, 10$^5$ &  4 -- 15000 &  1 -- 1285\\
\hline
            \noalign{\smallskip}
\end{tabular}
\tablefoottext{a}{\citet{Gusdorf082, Gusdorf11, Gusdorf12}}
\tablefoottext{b}{\citet{Anderl14}}

\end{table*}

The grid of shock models is that of G11, also used in other publications already mentioned (\citealt{Gusdorf12, Anderl14}). To summarise, we used a grid covering the following input parameters:~pre-shock density $n_{\rm H}$ from 10$^3$ to 10$^6$~cm$^{-3}$, magnetic field parameter from $b = 0.3$ to 2 (defined by $B$($\mu$ G) = $b \times [n_{\rm H}$ (cm$^{-3}$)]$^{1/2}$, where $B$ is the intensity of the magnetic field transverse to the shock direction of propagation), and shock velocity $\varv_{\rm s}$ from 15 to 35~km~s$^{-1}$. Our grid contains both stationary C- and J-type models, and also non-stationary, so-called CJ-type models (\citealt{Lesaffre041,Lesaffre042}). A more complete description of the parameters space coverage ($n_{\rm H}$, $b$, and $\varv_{\rm s}$) can be found in Table~\ref{table2}. The parameter coverage is not complete: not all velocities are present in our grid for all values of the magnetic-field parameter, $b$. In fact, the velocity of C-type shocks must remain below a critical value that depends mainly on the pre-shock density and magnetic-field parameter (\citealt{Lebourlot02,Flower031}), which explains the decrease of the maximum shock velocity with the pre-shock density in our grid. Additionally, the time necessary to reach a steadystate depends on the pre-shock density, shock velocity, and magnetic field values. Following the method presented in G08b (Section~4.1), the set of C-type shock models enabled us to restrict the range of the search in the parameter space for the CJ-type shock models. We then computed a grid of non-stationary shock models around a first estimate of the shock age, making sure that the range of ages was sufficient to include any model likely to fit the H$_2$ observational data. The final considered ages range from a few tens to around fifteen thousand years. Our grid also includes a few stationary models including the effects of grain--grain interactions, as presented in \citet{Guillet11}, \citet{Anderl13}, and already used in \citet{Anderl14}, \citet{Leurini142}.

We compared models and observations based on the three higher-$J$ CO lines, from CO (6--5) to (11--10), and on all the H$_2$ pure rotational lines. The CO (3--2) and (4--3) lines were excluded from our procedure, as stated above. Our initial CO (16--15) observation is in principle an upper limit, since it was associated with a slightly smaller beam than the other lines. Nevertheless we assumed that the CO (16--15) emission is as extended as the CO (6--5) emission, and we estimated the resulting flux over a 24$''$ beam by multiplying the value observed over a 16\farcs3 beam by a factor (16.3/24)$^2$. We then treated this corrected value as any other CO observation. Similar to what we did in previous studies, we used a $\chi^2$ routine to compare these observations to the models. The best results can be seen in figure~\ref{figure7}. On each of these panels, we show the points of our best fit in red circles, plus three other satisfying models in smaller, purple, green, and blue circles. Our best-fit model (in red points) is non-stationary, with the following characteristics:~$n_{\rm H} =10^4$~cm$^{-3}$, $b = 1.5$, $\varv_{\rm s} = 22$~km~s$^{-1}$, and an age of 3800 years. The $\chi^2$ value of our 10 and 20 best-fit models are within a factor 1.4 and 2.1 to the best (lowest) value. Broadly speaking, we found the H$_2$ lines in particular very difficult to fit: the exact curvature of the excitation diagram is only approached by our best model. We tried in vain to improve the quality of the fit by changing the ortho-to-para ratio of H$_2$ in our calculations (switching its value from 3 to 1 and 2). Eventually, we can not exclude that a better fit could be found via a better gridding of our parameters space. However, it turned out that this best-fit model also fitted the CO (3--2) and (4--3) lines quite in a very satisfying way. These values can be compared to what was constrained by G11: $n_{\rm H} =10^4$~cm$^{-3}$, $b$ = 1 -- 2, $\varv_{\rm s} = 25-30$~km~s$^{-1}$, and an age of 300-800 years. A few comments can be made on these shock parameters.

\textit{Shock type}. The shock type we constrained is the same as found by G11. Indeed, it is a very general result that in low-mass star forming environments, H$_2$ excitation diagrams can be fitted by these kinds of models only (e.g. \citealt{Giannini06, Gusdorf082, Flower13}). All of our ten best fits are CJ-type models. Even more, the $J$- and $C$-type models are all at the end of our list of best-fit models: the $C$-type models generally do not fit the pure rotational H$_2$ excitation diagram, while the $J$-type models do not generate enough CO emission to match the observations.

\textit{Pre-shock density}. First, the pre-shock density value is the same as that found in previous analyses of BHR71 (\citealt{Giannini04} for the analysis of the HH320 region, G11 for the SiO knot position), and in the resembling pure shock position L1157-B1 (e.g. \citealt{Gusdorf082}, \citealt{Flower12}). It is also the value associated with our ten best fits to the dataset; it is relatively well constrained, as in particular the general shape of the CO diagram is very sensitive to the density, i.e. to the pre-shock density parameter. In our ranking of models by decreasing $\chi^2$ value, the first model with a different pre-shock density has a $\chi^2$ value that is more than five times the lowest one. Generally speaking, it is a very standard value when modelling shocks from low-mass YSOs (also see HH54, \citealt{Giannini06}), which was also found in shocks associated with SNRs that are interacting with the interstellar medium (\citealt{Cesarsky99, Gusdorf12, Anderl14}). This corresponds to a post-shock density of $\sim$10$^5$~cm$^{-3}$, a value that is one or two orders of magnitude smaller than those found by Gia11 in the HH320 and 321 regions of the same outflow. This could be due to the fact that their study only focussed on the brightest pixel associated with these regions, whereas we consider a rather large beam size. On the other hand, our value is just above the H$_2$ density averaged over the whole inner outflow found by N09 (10$^{3.8}$~cm$^{-3}$). This can be explained by the fact we are focussing on a bright H$_2$ spot, and also that the post-shock value we are providing here is that of a stationary post-shock. In the (realistic) case where the shock has a finite spatial size, the post-shock gas will expand and come into pressure equilibrium with its surroundings, resulting in a globally lower value than ours. 

\textit{Magnetic field strength.} The strength of the transverse magnetic field is 150~$\mu$G in the pre-shock region, that turns into 1.62~mG in the post-shock, given the compression factor and the fact that the magnetic field is frozen in the neutral fluid. It is comparable to what was constrained in G11 ($b$ = 1--2), and also to what was found in the studies of HH320 in BHR71, L1157-B1, or HH54, or in the aforementioned SNR studies. This value is also consistent with the analysis of Zeeman measurements in molecular clouds where magnetic and kinetic energy densities are in approximate equipartition \citep{Crutcher99}. Our ten best fits all predict a value of the $b$ parameter between 1.0 and 2.0. Figure~\ref{figure7} shows the best model with a $b$ value outside of this range, in purple points. This model is associated to a $\chi^2$ value that is 1.8 times the lowest. This model is also a bit older than our ten best-fit ones (see below). It does not fit the lower-$J$ data as satisfyingly as our best-fit model (red points in Figure~\ref{figure7}).

\textit{Shock velocity.} The shock velocity values of our ten best-fit models all are between 20 and 25~km~s$^{-1}$. In our ranking of models by decreasing $\chi^2$ value, the first model with a velocity out of this range has a $\chi^2$ value that is about three times the lowest one. More specifically, the shock velocity that we constrain is slower than, e.g. the full-width at zero intensity of the CO lines. The modelled shock velocity is the difference between the pre-shock velocity and the impact velocity. Consequently, if the pre-shock material is already accelerated, the actual shock velocity is expected to be correspondingly lower.  Preliminary acceleration of the pre-shock is possible in the SiO knot position, as it has already been encompassed by the shock that is now propagating at the northern tip of the outflow (a bit more than twice as far from the driving protostars). Alternatively or simultaneously, this velocity discrepancy could be the sign of a limitation to our models. Indeed, we are modelling a multi-dimensional complex shock region, which is seen edge-on, by means of a one-dimension model, which is considered to be face-on. Additionally, it is likely that the rather large beam of our observations contains not one shock structure, but a collection of them. This can be seen in the spatial sub-structure of the H$_2$ emission revealed in figure~\ref{figure6}, where multiple peaks are detected. This can also be seen in the CO (figures~\ref{figure3} and~\ref{figure5}) or SiO (G11) line profiles, where a \lq plateau' or a \lq bump' can be seen between 20 and 40~km~s$^{-1}$, very much resembling the bullets revealed in the CO (in e.g. Cep E, \citealt{Gomezruiz12}) or water line profiles (in e.g. L1448-mm, \citealt{Kristensen11}). Another expression of the intrinsic multi-dimensional nature of the observed shocks is the impossibility of fitting the different molecules with the same filling factor values. For instance, H$_2$ is only associated with the hottest regions in our beam, as a 500~K temperature is necessary to populate its levels, whereas CO is more easily excited (for $J_{\rm up}$ = 6, $E_{\rm up}$  = 116~K). In particular, unlike H$_2$, CO is probably emitting in the post-shock expansion region mentioned in the \lq pre-shock density' paragraph above, where the primary shock structure progressively mixes with the ambient medium in all spatial dimensions. This effect cannot be accounted for by a 1-D shock model.

\textit{Shock age.} Finally the shock age is rather large: it is larger than what was previously constrained for this region of BHR71 by G11 (300--800 years), which is due to the high amounts of CO that we detected, that are not compatible with younger CJ-type shocks. However, again the age predicted by our ten best-fit models consistently lies between 3500 and 5500 yr, with the model in purple in the Figure~\ref{figure7} being slightly older. In our ranking of models by decreasing $\chi^2$ value, the first model with an age younger than 3000 years has a $\chi^2$ value that is about twice the lowest. The age that we find is between the dynamical age of the SiO knot and that of the global northern lobe (see Section~\ref{sub:daoto}), which hints again at the fact that we are probably catching several shock episodes in our beam.

\textit{Limitations.} The question of the largest source of uncertainty in these results is difficult to assess. The observational problems (the different beam sizes and the fact that the SiO knot lies close to the edge of the H$_2$ observations) make for an important limitation, but we believe we have taken it into account by using conservative errorbars. More problematic is the complex nature of shocks structures. Lacking high angular resolution, we could only assume that we catch a single shock structure in our beam. Our results seem to indicate this is not the case, given the constrained shock velocity and ages, and the different filling factors we had to adopt for each molecule. A shock similar to those we have in our grid of models is probably propagating in one beam of 24$''$, but it is very likely to be associated with less violent shocks corresponding to the mixing of its warm, dense, and accelerated post-shock with the surrounding ISM. To summarise, we could only be aware and accept the fact that we are averaging processes out by working over a beam size typical of single-dish observations. From the modelling point of view, solutions do exist to more realistically account for the observations of complex geometrical shock structures. The first consists of adopting a probability density function (PDF) of shocks, which is to consider that the observed shock characteristics follow a statistical distribution. This distribution can unfortunately only be guessed, and hence generates additional free parameters (see \citealt{Lesaffre13} for applications of this method). The second consists of stitching 1-D shock layers (similar to the one we consider here) to either a curve (e.g. \citealt{Kristensen08}) or a bow-shock structure (e.g. \citealt{Gustafsson10}), hence simulating \lq pseudo-' 2- or 3-D shock propagation. This approach also yields free parameters, such as the inclination of the magnetic field with respect to the shock structure. None of these methods would be ideal in the present case, given the lack of knowledge of the small-scale shock structure within the 24$''$ beam (no indication on the collection of shocks is available at the moment), and the relatively limited dataset (if more free parameters are to be considered, more constraints are needed to lift the degeneracy in the results). Their application will become relevant when the SiO knot is observed at higher angular resolution (in CO and H$_2$), and if possible with a maximum number of observations in key species (such as H$_2$, CO, and SiO, but also H$_2$O, OH and \ion{O}{I} for instance). We note that at this point, a more tightly sampled shock model will prove necessary.

In the following we will present results based on our best fit only. All of our ten best-fit models were found to be non-stationary ($CJ$-type), with $n_{\rm H} =10^4$~cm$^{-3}$, $b = 1-2$, $\varv_{\rm s} = 20-25$~km~s$^{-1}$, and an age of 3500--4500 years.

\subsection{Employing shock models: SiO chemistry}
\label{sub:esmsioc}

   \begin{figure}
   \centering
   \includegraphics[width=9cm]{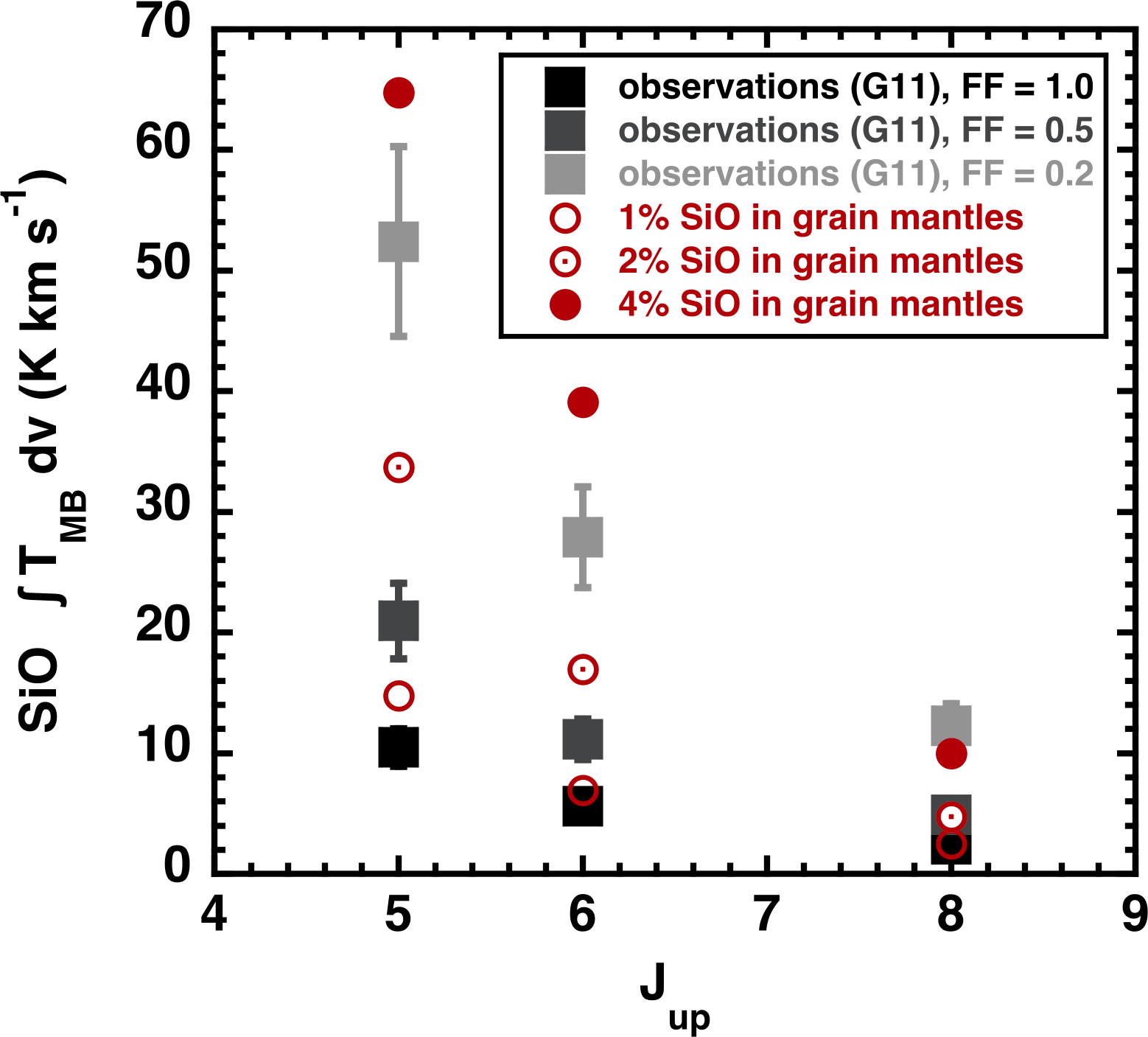}
      \caption{The integrated intensity diagram generated from the G11 observations of three SiO lines ($J_{\rm up}$ = 5, 6, 8), corrected for beam size and for filling factor effect: filling factor of 1, 0.5, and 0.2 in black, dark grey, and light grey squares. The result of our best-fit model for the H$_2$ and CO lines ($n_{\rm H} =10^4$~cm$^{-3}$, $b = 1.5$, $\varv_{\rm s} = 22$~km~s$^{-1}$, and an age of 3800 years) is shown with 1, 2, or 4\% of the pre-shock Si placed in the grain mantles in red empty, dotted, or filled circles.}
         \label{figure8}
   \end{figure}

After constraining shock models based on CO and H$_2$ observations, it is now possible to fine-tune the SiO chemistry. SiO has long been interpreted as a tracer of C-type shocks with a velocity greater than 20--25~km~s$^{-1}$ (e.g. \citealt{Caselli97,Schilke97,Gusdorf081}). Indeed, in these types of shocks, the drift velocity between the charged grains and the most abundant, neutral molecular species is sufficient to generate the sputtering of the core of the grains, where all the silicon-bearing material was considered to be locked. However \citet{Gusdorf082} have demonstrated: 1) that CJ-type shock models were the only ones to fit the H$_2$ emission in L1157-B1, and 2) that this process was not efficient enough to generate levels of SiO emission comparable to the observations in this kind of (CJ-type) shock models. In order to be able to self-consistently account for both SiO and H$_2$ emission, one solution was considered, consisting of transferring a fraction of SiO from the core to the mantle of the grains in the pre-shock phase. The maximum fraction of SiO thus transferred to the mantles was set to 10\%. With mantle sputtering being easier than core sputtering, these authors were then able to fit the H$_2$ and SiO emission by means of a single, non-stationary shock model. The presence of silicon-bearing material in the grain mantles has also been assumed in \citet{Coutens13} to explain the narrow emission component detected in the SiO (2--1) line profile at the systemic velocity around \mbox{NGC1333 IRAS 4A}.

Interestingly, the transfer of Si from the core to the mantles in the form of SiO was also considered by \citet{Anderl13} in the context of denser shock models. Indeed, for pre-shock densities above 10$^5$~cm$^{-3}$, the effect of grain--grain interactions cannot be neglected in shock models, as shown by \citet{Guillet11}. Taking these interactions into account results in the significant production of small, charged grains that couple very well with the neutral fluid, in effect reducing the width of the shock layer, and increasing its maximum temperature. In the narrow shock layers thus produced, \citet{Anderl13} have shown that the grain core sputtering is not efficient enough to produce levels of SiO emission comparable to the observations, hence the recourse to the transfer of a fraction of Si towards the mantles in the pre-shock phase. \citet{Leurini142} have validated this approach by successfully confronting these models with observations.
   
Consequently, we ran our best-fit model for the H$_2$ and CO data ($n_{\rm H} =10^4$~cm$^{-3}$, $b = 1.5$, $\varv_{\rm s} = 22$~km~s$^{-1}$, and an age of 3800 years), including 0 to 10\% of the pre-shock silicon-bearing material in the form of SiO in the grain mantles. We then generated the corresponding integrated intensity diagrams, which we compared with the observations. The result can be seen in figure~\ref{figure8}. For a filling factor value conservatively varied between 0.2 and 1, we show that no more than 4\% of SiO needs to be placed in the grain mantles to reproduce the observations. The presence of silicon-bearing material on the grain mantles could be explained by the fact that the SiO knot position has already been processed in the past by the shocks that are now located at the apex of the northern outflow lobe, further north, as can be seen in figure~\ref{figure1}. Under this assumption, we have demonstrated that it is possible to self-consistently fit H$_2$, and velocity-resolved CO and SiO observations in the \lq SiO knot', pure shock position of BHR71.  

\subsection{Employing shock models: energetics}
\label{sub:esme}

In this section, we discuss the energetics associated with the shocks along two axes: the CO excitation generated from our best-fit model, and the derivation of the outflow parameters.

\textit{CO excitation.} Indeed, the excitation of CO has been intensively used in the literature to demonstrate the presence of various processes at work in the regions of star formation. For example, \citet{Vankempen101} and \citet{Visser12} obtained CO flux diagrams (similar to our figure~\ref{figure7}) with PACS, centred on low-mass protostars at the origin of various outflows (HH46, NGC1333 IRAS 2A, and DK Cha). Based on these diagrams, they evidenced the existence of three physical components corresponding to three distinct processes contributing to the excitation of CO: a passively heated envelope, UV-heated outflow cavity walls, and small-scale shocks along the cavity walls. The SiO knot position is distant from the central protostars IRS 1 and IRS 2 (see e.g. figure~\ref{figure1}), which most likely prevents any UV heating from the central star to be operating. Furthermore, it lies outside the envelope, as revealed by IRAC \citep{Tobin10}, NH$_3$ (1,1) \citep {Bourke95,Bourke97} or N$_2$H$^+$ \citep{Chen08} emission. In other words, The SiO knot position offers the opportunity to study pure small-scale shocks along cavity walls. However, the comparison of our shock models with those presented in \citet{Vankempen101} or \citet{Visser12} is not really meaningful. These authors indeed used a multiple-shock layers model, with a pre-shock density of 10$^{6.5}$~cm$^{-3}$ close to the protostar, and 10$^{4.5}$~cm$^{-3}$ further away along the envelope (see figure 4 of \citealt{Visser12}). Furthermore the 1D shocked layers they used were outputs of the Paris-Durham model we are also using in the present study. However they did not include the tip of the envelope that would correspond to the SiO knot position, as they solely focus on the outflow cavity in the vicinity of the protostar. The conclusion from our analysis is that our constrained pre-shock density of 10$^4$~cm$^{-3}$ is consistent with the range of values they used (as they considered a decreasing density further away from the protostar).

   \begin{figure}
   \centering
   \includegraphics[width=9cm]{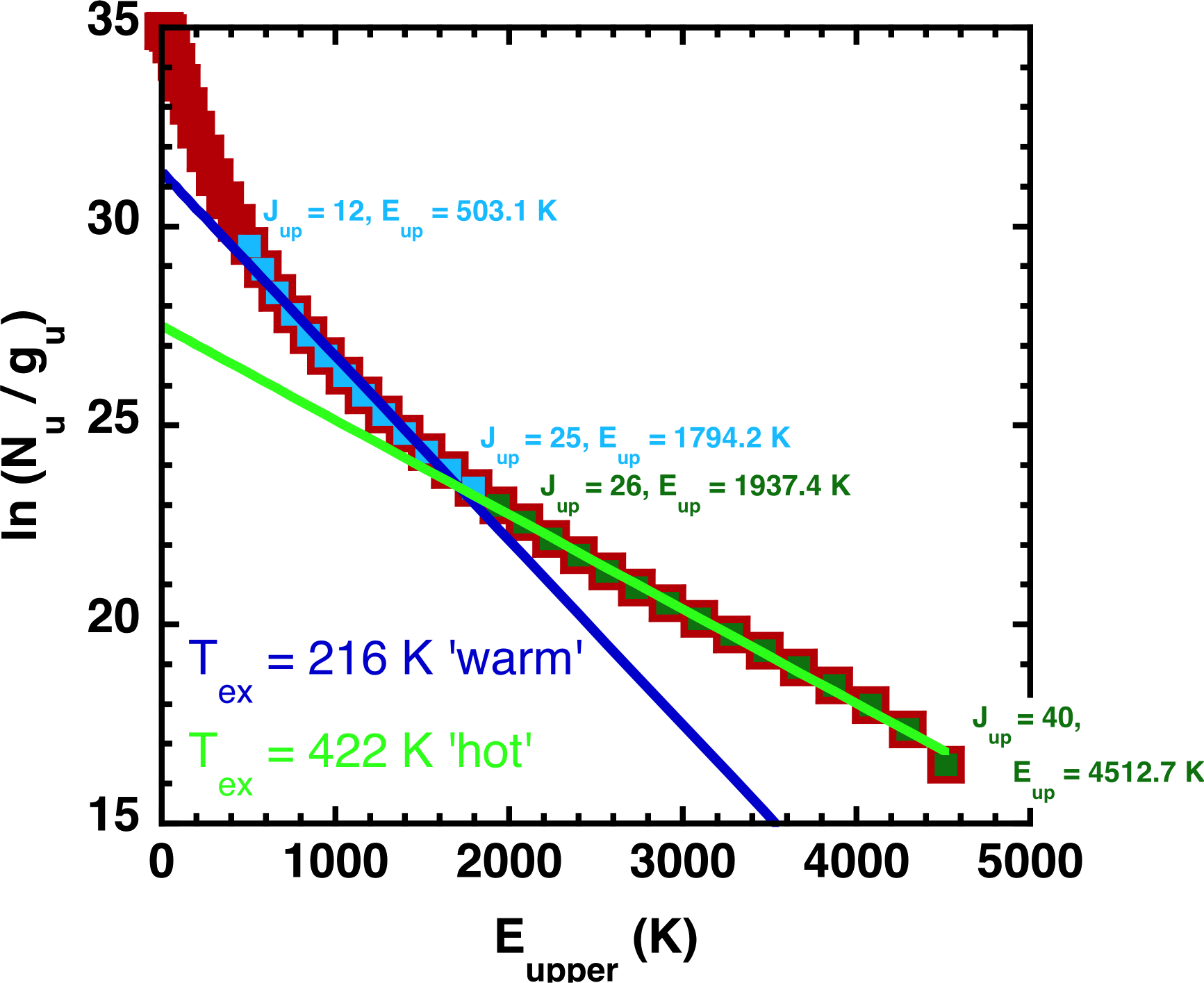}
      \caption{The CO excitation diagram produced from our best-fit model (red squares). Two temperature components can be fitted to our shock model over the PACS range of observations:~ a warm component ($T_{\rm ex}$ = 216~K, dark blue line) fitting the level populations for $J_{\rm up}$= 12 to 25 (light blue squares), and a hot component (\mbox{$T_{\rm ex}$ = 422~K}, light green line) fitting the level populations for $J_{\rm up}$= 26 to 40 (dark green squares).}
         \label{figure9}
   \end{figure}

Another, more classical approach to discuss energetics and physical conditions based on CO excitation consists of using excitation diagrams, as reviewed in \citet{Visser14}. Figure~\ref{figure9} presents the excitation diagram built from our best-fit model for CO. As extensively described in \citet{Goldsmith99}, in local thermodynamical equilibrium conditions and under optically thin regime, the points in this type of diagram are expected to fall on a straight line, whose inverse of the slope is the excitation temperature of the transitions (also see the description of H$_2$ excitation diagrams in section~\ref{sub:h2o}). As pointed out in \citet{Visser14}, numerous studies can be found in the literature describing the building of this kind of excitation diagrams based on PACS observations centred on outflow-driving protostars of low- to high-masses, and their subsequent decomposition in at least two gas components, one warm ($T_{\rm ex}$=320$\pm$50~K), and one hot ($T_{\rm ex}$=820$\pm$150~K). Usually the breakpoint between these two components is around 1800~K, between the (25--24) and (26--25) transitions. We have produced a similar diagram from our best-fit model, as can be seen in figure~\ref{figure9}. For comparison purposes, we have also applied a two-component linear fit of the lines accessible to PACS, with the warm component ($T_{\rm ex}$ = 216~K) fitting the level populations for $J_{\rm up}$= 12 to 25, and a hot component ($T_{\rm ex}$ = 422~K) fitting the level populations for $J_{\rm up}$= 26 to 40. The values we constrained for the excitation temperatures within these two components are systematically below those inferred from PACS observations centred on outflow-driving protostars of all possible mass \citep{Vankempen102,Herczeg12,Goicoechea12,Manoj13,Karska13, Karska141, Green132, Green131, Dionatos13,Lee13,Lindberg14}. This shows that pure shocks contribute to both the so-called \lq warm' and 'hot' components. Although the excitation temperature associated with the warm component is close to the observed values, the figure also shows that pure shocks fail to entirely account for any of these warm or hot components. Indeed the excitation temperatures that we find here are less than those derived from the observations in those papers. This means that close to the protostar, additional mechanisms, or different kinds of shocks should account for the CO observations. It also shows to which extent the collection of continuous temperature components that is generated by our shock models can be interpreted as a two-excitation component, through the analysis of CO excitation diagrams, for transitions with $J_{\rm up}$=12 to 40.

\textit{Outflow parameters.} Finally, we studied the energetics associated with this pure shock position. Our first method is to follow the procedure presented in \citet{Anderl14} for the W44 SNR shock study. Indeed, we can infer the mass, momentum, and energy associated with our best-fit model, under the assumption of a filling factor equal to one. The results are presented in the second column of table~\ref{table3}. First, the total mass contained in the beam is $1.8\times10^{-2}~M_\odot$. This value is far greater than that determined for the mapped area (half of the inner part of the outflow) by N09 (2.5$\times 10^{-3} M_\odot$), or Gia11 (0.6$\times 10^{-2} M_\odot$). This might be partly explained by the fact that our beam size is rather large and that we decided to focus on the brightest H$_2$ spot in the map, contrary to those studies that operated on values averaged over larger areas. More convincingly, this is probably due to the fact that our observations--models approach gives us access to the population of the ($v$ = 0, $J$ = 0, 1) H$_2$ levels, contrary to the observations. The contribution of these two levels to the total column density is significant: a linear fit to the \textit{Spitzer} observations presented in Figure~\ref{figure7} yields a column density $N$(H$_2$)$\sim 6.9\times 10^{19}$~cm$^{-2}$, whereas linearly fitting the ($v$ = 0, $J$ = 0, 1) part of the modelled rotational diagram yields $N$(H$_2$)$\sim 1.8\times 10^{21}$~cm$^{-2}$, which is 26 times larger than that measured based only on the observations. Moreover, our value is consistent with the total mass of 1$M_\odot$ for the northern lobe determined by \citet{Bourke97} based on CO (1--0) and (2--1) line observations (combined with $^{13}$CO and C$^{18}$O (1--0)). From a different perspective, the value that we found is simultaneously $\sim$50 times smaller than that found in the bright CO positions of the W44 SNR studied in the same way by \citet{Anderl14}, where molecular shocks also propagate. The corresponding momentum is $0.4~M_\odot$~km~s$^{-1}$, also a good factor $\sim$100 below the values inferred using similar methods in W44. This is also a factor 10 below the value found based on more observational methods (described in the next paragraph) over a 12\farcs5 beam in the massive star-forming region W28~A2 that encompasses three different outflows at least (Gusdorf et al., in press). Finally, the dissipated energy is $4.2 \times 10^{43}$~erg, typically two orders of magnitudes below the equivalent quantity in W44. This is also two orders of magnitudes below the energy dissipated in a 12\farcs5 beam in W28~A2 but comparable to what was found by \citet{Gomezruiz13} for the entire outflows (associated with low-mass YSOs) L1448-IRS3 and HH211-mm (based on the use of more observational methods). From this method, it seems that BHR71 can be defined as an energetic low-mass outflow, but the relatively high numbers we found could be the effect of the method we used, based on a sophisticated shock model. From the energetic point of view, the impact of the whole BHR71 outflow on the Galactic ISM is much less than that, e.g. of the whole W44 SNR, because the dissipated energy is smaller, but also because of its much smaller size: the BHR71 outflow is roughly covered by a rectangle of $\sim 0.1\times0.5$~pc$^2$, whereas a SNR such as W44 resembles a circle of $\sim$26~pc radius.

We also used a second method, following \citet{Bontemps96,Beuther02}, to calculate the outflow parameters related to the considered species (CO, H$_2$ and SiO). These parameters (dynamical age $t_{\rm d}$, mass $M$, mass entrainment rate $\dot{M}$, momentum $P$, mechanical force $F_{\rm m}$, kinetic energy $E_{\rm k}$, and mechanical luminosity $L_{\rm mech}$, are calculated using the equations:
\begin{equation}
t_{\rm d} = R/ \delta \varv_{\rm max}, \\
\label{equation2}
\end{equation}
\begin{equation}
M = N \times \pi R^{2} \times m_{\rm species}, \\
\label{equation3}
\end{equation}
\begin{equation}
\dot{M} = M / t_{\rm d},
\label{equation4}
\end{equation}
\begin{equation}
P = M \times \delta \varv_{\rm max},
\label{equation5}
\end{equation}
\begin{equation}
F_{\rm m} = M \times \delta \varv_{\rm max} / t_{\rm d}, 
\label{equation6}
\end{equation}
\begin{equation}
E_{\rm k} = M \times \delta \varv_{\rm max}^2 / 2,
\label{equation7}
\end{equation}
\begin{equation}
L_{\rm mech} = E_{\rm k} / t_{\rm d},
\label{equation8}
\end{equation}
where $N$ is the column density of the considered species, $R$ the radius of our analysis region, and $\varv_{\rm max}$ the approximate zero-base linewidth of the considered species. We used CO, H$_2$, and SiO column densities extracted from our best-fit model associated with a filling factor of 1. The results are indicated in table~\ref{table3}. Lacking the necessary spectral resolution for the H$_2$ data, we assumed that the $\delta \varv_{\rm max}$  and $t_{\rm d}$ values for H$_2$ are identical identical as for CO. This leads to overestimating the mass, momentum, and energy calculated based on the H$_2$ data with respect to the results purely obtained from the model (first method). This is because the global shock velocity of our best-fit model, 22~km~s$^{-1}$, is less than the zero base linewidth that we attributed to H$_2$ in this second method. Regardless, the CO mechanical luminosity we constrain is again consistent with that measured by \citet{Bourke97} for the whole outflow (0.5~$L_\odot$), whereas the corresponding value we calculated for H$_2$ exceeds that given by N09 for the fraction of the outflow that they mapped (0.05~$L_\odot$).

\begin{table}
\caption{Outflow parameters over the beam of our observations in H$_2$, CO, and SiO. The $\delta \varv_{\rm max}$  and $t_{\rm d}$ values for H$_2$ were assumed to be identical as for CO. The second column summarises the values extracted from the best-fit model (first method, see text).}             
\label{table3}      
\centering                          
\begin{tabular}{l  c c  c  c}        
\hline           
Species & Model & H$_2$ & CO & SiO \\
\hline
\hline
$N$ (cm$^{-2}$) & 3.7e21 & 1.9e21 & 3.2e17 & 2.0e15 \\
$M$ ($M_\odot$) & 1.8e-2 & 1.3e-2 & 1.5e-5 & 1.5e-7 \\
$\delta \varv_{\rm max}$ (km~s$^{-1}$) & -- & 40 & 40 & 40 \\
$t_{\rm d}$ (yr) & -- & 1785 & 1785 & 1785 \\
$\dot{M}$ ($M_\odot$ yr$^{-1}$) & -- & 7.0e-6 &8.6e-9 & 8.4e-11 \\
$P$ ($M_\odot$ km s$^{-1}$) & 0.4 & 5.0e-1 & 6.1e-4 & 6.0e-6 \\
$F_{\rm m}$ ($M_\odot$ km s$^{-1}$ yr$^{-1}$) & -- & 2.8e-4 & 3.4e-7 & 3.4e-9 \\
$E_{\rm k}$ (erg) & 4.2e44 & 2.0e44 & 2.4e41 & 2.4e39 \\
$L_{\rm mech}$ ($L_\odot$) & -- & 9.3e-1 & 1.1e-3 & 1.1e-5 \\
\hline
\hline
\end{tabular}
\end{table}

We can now compare the outflow parameters as inferred from our observations of BHR71 with similar studies found in the literature for outflows associated with protostars of various masses. An impressive number of such studies have been aimed at isolated targets, making it difficult to give some perspective on our results based on a one-to-one comparison (e.g. \citealt{Leurini06}, \citealt{Lebron06}, \citealt{Fontani09}, \citealt{Liu10}, \citealt{Guzman11}, \citealt{Cyganowski11}). We consequently focus on a comparison with \lq survey studies', i.e. studies that are aimed at extracting outflow parameters from a sample of sources. From this respect, we found that the outflow parameters derived from CO observations of BHR71 are obviously systematically less than the values inferred in more massive environments (e.g. \citealt{Duartecabral13, Klaassen11}), let alone in the outflows associated with O-type stars \citep{Lopezsepulcre09}. This is also true for SiO-related energetic parameters: BHR71 is less energetic from this point of view than its massive counterparts \citep{Sanchezmonge13}. Indeed these authors computed the mass, momentum, and energy associated with the SiO (2--1) and (5--4) lines. Their study was focussed on the whole outflows, resulting in values considerably larger than those inferred here (4-110 and 2-35 $M_\odot$; 26-2130 and 11-440$M_\odot$ km s$^{-1}$; (0.2-75) and (0.06-16)$\times 10^{46}$ergs). On the contrary, the energetic parameters inferred from our study of one bright beam are similar to those inferred over the extent of the whole outflows associated with similar low-mass stars as IRS1, for instance. This is the case for the outflows in the Perseus cloud (NGC1333, L1448, HH211, L1455, e.g. \citealt{Curtis10, Gomezruiz13}). The conclusion is that either BHR71 is indeed more energetic than its low-mass counterparts, or that our method partly based on shock modelling naturally yields  higher numbers than in all these studies, that often make use of fewer CO lines, for instance, than in the present work.

\section{Conclusions}
\label{sec:conc}

We have presented new observations of the BHR71 bipolar outflows, obtained with SOFIA in $^{12}$CO (11--10), and (16--15), and with APEX in $^{12}$CO (3--2), (4--3), (6--5), (7--6), and $^{13}$CO (3--2).

We combined these data with existing datasets: in H$_2$ (\textit{Spitzer}/IRS) and SiO (APEX), and we have compared the observations in the form of a CO flux diagram, an H$_2$ excitation diagram, and an SiO integrated intensity diagram to a grid of sophisticated shock models.

Our best fit is non-stationary ($CJ$-type) and has the following parameters: $n_{\rm H} =10^4$~cm$^{-3}$, $b = 1.5$, $\varv_{\rm s} = 22$~km~s$^{-1}$, and an age of 3800 years.  The age and velocity of the shock model hint at the presence of more than one shock structure within the rather large beam of our observations, 24$''$. This was also suggested by the fact that we had to assume different filling factor values for the different considered species. From the analysis of our ten best-fit models, we consider that the constrained values are quite robust, as these models all had $n_{\rm H} =10^4$~cm$^{-3}$, $b = 1-2$, $\varv_{\rm s} = 20-25$~km~s$^{-1}$, and an age of 3500--4500 years.

However this modelling can still be used to discuss the feedbacks of the shocks encompassed in our observations. We studied its chemical feedback in terms of SiO chemistry, placing an upper limit of 4\% of the total silicon-bearing material in the form of SiO in the grain mantles in the pre-shock region.

We quantified the contribution of shocks to the excitation of CO around low-mass protostars surrounded by outflows, and shown that the CO excitation diagram from a shocked layer where the gas temperature is calculated point by point can be interpreted as a two-component one for levels from $J_{\rm up}$=12 to 40.

We inferred global outflow parameters from our shock model: a mass of $1.8\times10^{-2}~M_\odot$ was measured in our beam, in which a momentum of $0.4~M_\odot$~km~s$^{-1}$ was dissipated, corresponding to an energy of $4.2 \times 10^{43}$~erg. We also analysed the energetics of the outflow species by species. Both methods suggest that BHR71 is a rather energetic outflow.

Three perspectives lie ahead of the present study. The first is to generalise our analysis to the whole outflow, and to observationally constrain the outflow energetic parameters over the whole mapped area. This will yield meaningful comparisons with observational studies of various similar objects. The second is to observe the SiO knot position with ALMA to resolve the multiple shock structures caught in our present single-dish beam. This should allow us to isolate a single shock structure, which would help us to get closer to the geometrical configuration that the Paris-Durham model was tailored to fit, and lead the shock analysis to a new level of understanding, both in terms of chemistry and energetics. Finally, at this point, more emission lines will also have to be included in our analysis, especially H$_2$O lines.

\begin{acknowledgements}
      We thank an anonymous referee for comments that helped to improve the quality of this work. We thank the SOFIA operations and the GREAT instrument teams, whose support has been essential for the GREAT accomplishments, and the DSI telescope engineering team. Based [in part] on observations made with the NASA/DLR Stratospheric Observatory for Infrared Astronomy. SOFIA Science Mission Operations are conducted jointly by the Universities Space Research Association, Inc., under NASA contract NAS2-97001, and the Deutsches SOFIA Institut, under DLR contract 50 OK 0901. This work was partly funded by grant ANR-09- BLAN-0231-01 from the French Agence Nationale de la Recherche as part of the SCHISM project. It was also partly supported by the CNRS programme \lq Physique et Chimie du Milieu Interstellaire'.    
 \end{acknowledgements}

%-------------------------------------------------------------------

\bibliographystyle{aa}
\bibliography{biblio}

\Online
\begin{appendix} 
\section{Filling factors}   
\label{sec:ff}   

\begin{figure}
   \centering
   \includegraphics[width=9cm]{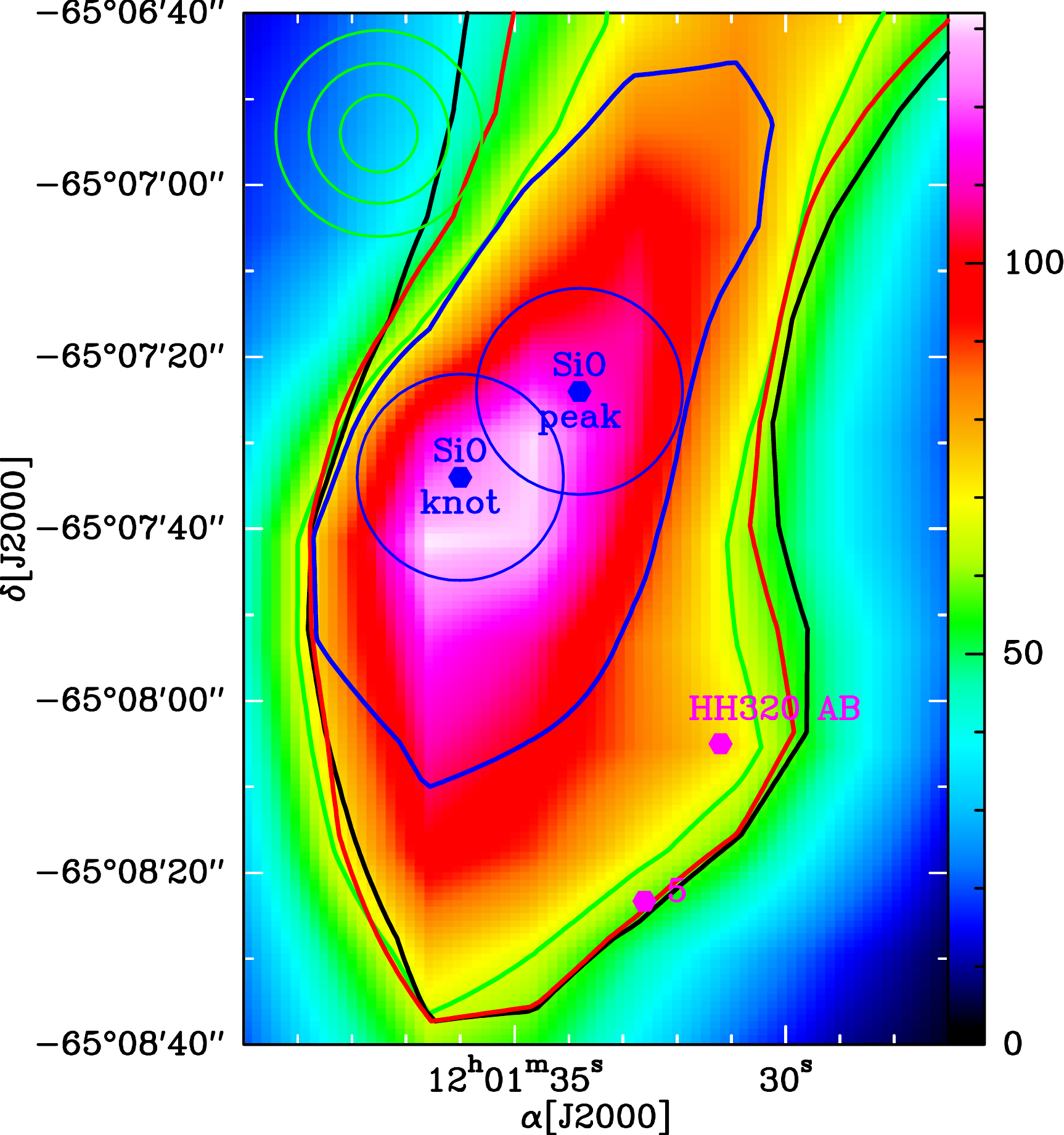}
      \caption{The same field as in figure~\ref{figure2}, shown in CO (6--5) (colours) convolved at the 24$''$ resolution. The half-maximum contours of the emission of the CO (3--2) (black line), (4--3) (red), (6--5) (green), and (7--6) (blue) lines is also shown at the same 24$''$ resolution. The SiO peak and knot, HH320 AB and knot 5 positions, as well as the beam sizes of the (6--5), (16--15) and (11--10) transitions are also indicated as in figure~\ref{figure2}.}
         \label{figurea1}
   \end{figure}

The half-maximum emission contours of the CO (6--5) and (3--2) lines were already shown in figure~\ref{figure2}, in thick black and red contours, at their nominal resolutions of 9$''$ and 18\farcs1. For both the SiO peak and knot, the filling factor of the emission with respect to a beam of 24$''$ (that of the CO (11--10) line observations), inferred from the red, CO (3--2) contours is of the order of 1. As we performed our analysis over a 24$''$ beam size, and not at the resolution of the CO (3--2) line, we decided to verify that this filling factor assumption was correct for all CO transitions at a 24$''$ resolution. We hence convolved all the APEX data to this resolution. The result is shown in figure~\ref{figurea1}. The dataset is remarkably consistent in terms of the size of the emitting region:~the half-maximum emission contour for all lines is broadly the same, except for the (7--6) transition, which could be due to insufficient signal-to-noise values. Based on this figure, we chose to operate with a filling factor of 1 for all CO transitions observed in the SiO knot position over a beam of 24$''$.

\section{CO, SiO, and H$_2$ observables}
\label{sec:csh2o}

\begin{table}
\caption{CO and SiO integrated intensity values, and H$_2$ level populations measured over a beam of 24$''$ centred on the SiO knot position. This means that the CO (16--15) (nominal resolution of 16\farcs3) and SiO (nominal resolution of 28$''$) data were \textit{a posteriori} corrected assuming the emission is extended over the largest beam. As such, the initial data were multiplied by (16.3/24)$^2$ and (28/24)$^2$. The uncertainties are given. The integrated intensities were calculated between -4.5 and 50~km~s$^{-1}$. The CO and SiO integrated intensities correspond to a filling factor of 1.}             
\label{tablea1}      
\centering                          
\begin{tabular}{l  c c  c }        
\hline           
Observable & Species & Line & SiO knot  \\
\hline
$\int T_{\rm MB} d\varv$ (K km s$^{-1}$) 	& CO & (3--2) 	  & 113.4$\pm$17.0  \\
					& CO & (4--3) 	  & 117.3$\pm$17.6 \\
					& CO & (6--5) 	  & 115.2$\pm$17.3  \\
					& CO & (7--6)	  & 95.2$\pm$14.3    \\
					& CO & (11--10) & 22.9$\pm$3.4      \\
					& CO & (16--15) & 3.4$\pm$0.5  \\
$\int T_{\rm MB} d\varv$ (K km s$^{-1}$) 	& SiO & (5--4) 	  & 10.5$\pm$1.1  \\
					& SiO & (6--5) 	  & 5.6$\pm$0.6 \\
					& SiO & (8--7) 	  & 2.5$\pm$0.3  \\
\hline
Observable & Species & Level & SiO knot  \\
\hline
ln (N/g) 	& H$_2$ & (0,0) & 43.5$\pm$1.2  \\
 		& H$_2$ & (0,1) & 41.0$\pm$1.3  \\
 		& H$_2$ & (0,2) & 40.3$\pm$1.3  \\
		& H$_2$ & (0,3) & 38.9$\pm$1.4  \\
		& H$_2$ & (0,4) & 38.2$\pm$1.4  \\
 		& H$_2$ & (0,5) & 36.7$\pm$1.4  \\
 		& H$_2$ & (0,6) & 35.7$\pm$1.4  \\
 		& H$_2$ & (0,7) & 34.6$\pm$1.5  \\
\hline
\hline
\end{tabular}
\end{table}

\section{The CO flux diagram in the SiO peak position}
\label{sec:tcofditspp}

   \begin{figure}
   \centering
   \includegraphics[width=9cm]{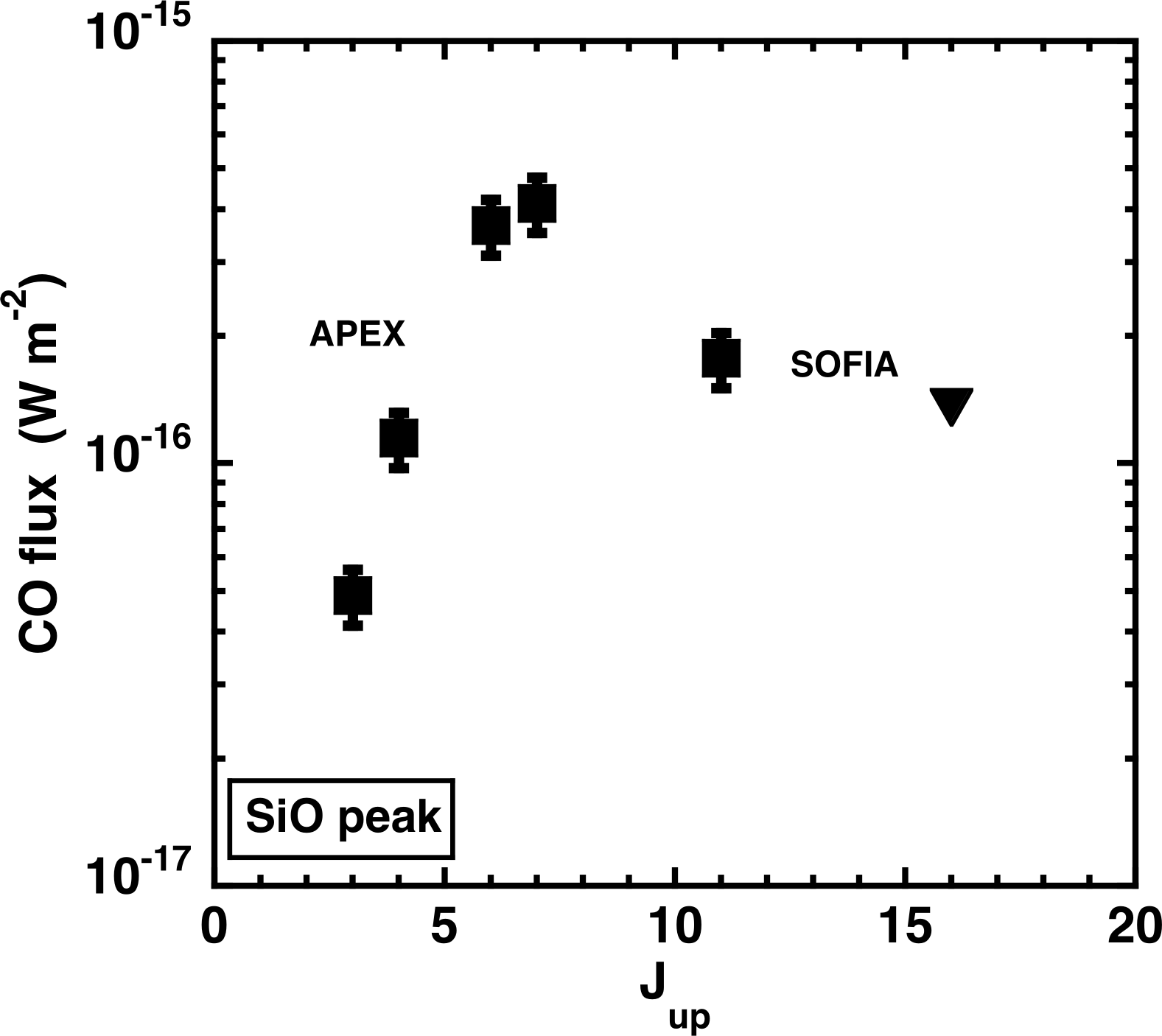}
      \caption{The observational CO flux diagram over a beam of 24$''$ obtained in the SiO peak position. The CO (16--15) line is not detected: an upper limit estimate is displayed in the form of a black arrow.}
         \label{figurec1}
   \end{figure}

\end{appendix}

\end{document}